\documentclass[twocolumn]{aastex62}
\usepackage{amsmath}
\usepackage{here}

\newcommand{\ring}{\mathrm{ring}}
\newcommand{\gas}{\mathrm{g}}
\newcommand{\dst}{\mathrm{d}}
\newcommand{\drift}{\mathrm{drift}}
\newcommand{\sigmad}{\Sigma_{\dst}}

\newcommand{\tstop}{t_{\mathrm{stop}}}

\newcommand{\cs}{c_{\mathrm{s}}}
\newcommand{\cd}{c_{\mathrm{d}}}
\newcommand{\hd}{H_{\mathrm{d}}}

\def\msun{M_{\odot}}
\def\merth{M_{\oplus}}

\received{--}
\accepted{--}
\submitjournal{ApJ}

\shorttitle{Nonlinear growth of secular gravitational instability}
\shortauthors{Tominaga et al.}

\begin{document}

\title{Secular Gravitational Instability of Drifting Dust in Protoplanetary Disks: \\Formation of Dusty Rings without Significant Gas Substructures}

\correspondingauthor{Ryosuke T. Tominaga}
\email{tominaga.ryosuke@a.mbox.nagoya-u.ac.jp}

\author[0000-0002-8596-3505]{Ryosuke T. Tominaga}
\affil{Department of Physics, Nagoya University, Nagoya, Aichi 464-8692, Japan}

\author{Sanemichi Z. Takahashi}
\affiliation{National Astronomical Observatory of Japan, Osawa, Mitaka, Tokyo 181-8588, Japan}

\author{Shu-ichiro Inutsuka}
\affiliation{Department of Physics, Nagoya University, Nagoya, Aichi 464-8692, Japan}



\begin{abstract}
Secular gravitational instability (GI) is one of the promising mechanisms for creating annular substructures and planetesimals in protoplanetary disks. We perform numerical simulations of the secular GI in a radially extended disk with inward drifting dust grains. The results show that, even in the presence of the dust diffusion, the dust rings form via the secular GI while the dust grains are moving inward, and the dust surface density increases by a factor of ten. Once the secular GI develops into a nonlinear regime, the total mass of the resultant rings can be a significant fraction of the dust disk mass. In this way, a large amount of drifting dust grains can be collected in the dusty rings and stored for planetesimal formation. In contrast to the emergence of remarkable dust substructures, the secular GI does not create significant gas substructures. This result indicates that observations of a gas density profile near the disk midplane enable us to distinguish the mechanisms for creating the annular substructures in the observed disks. The resultant rings start decaying once they enter the inner region stable to the secular GI. Since the ring-gap contrast smoothly decreases, it seems possible that the rings are observed even in the stable region. We also discuss the likely outcome of the non-linear growth and indicate the possibility that a significantly developed region of the secular GI may appear as a gap-like substructure in dust continuum emission since dust growth into larger solid bodies and planetesimal formation reduce the total emissivity.
\end{abstract}



\section{Introduction}\label{sec:intro}
A protoplanetary disk forms around a newborn star and is the site of planetesimal and planet formation. Recent ALMA observations have been exploring disk structures at high angular resolutions. One of the intriguing reports is that many Class II disks have single or multiple dust rings/gaps at various radii $\gtrsim10$ au \citep[e.g.,][]{ALMA-Partnership2015,Andrews2016,Tsukagoshi2016,Isella2016,Fedele2017,Long2018,Andrews2018}. The rings and gaps have been observed not only in disks with the age of a few - 10 Myr (e.g., TW Hya, \citealt{Andrews2016,Tsukagoshi2016}; HD169142 \citealt{Fedele2017,Perez2019a}; CI Tau, \citealt{Clarke2018}; Sz 91, \citealt{Tsukagoshi2019}) but also in very young ($\le$ 1 Myr) disks (e.g., HL Tau, \citealt{ALMA-Partnership2015}; WL 17, \citealt{Sheehan2017}; Elias 42, \citealt{Dipierro2018}; GY 91, \citealt{Sheehan2018}). 
The origin of those annular substructures is still under debate. 

One possible scenario is that (sub-)Jupiter mass planets already exist and carve gaps \citep[e.g.,][]{Gonzalez2015,Kanagawa2015,Zhang2018}. 
Although recent work reported kinematic signatures of Jupiter-mass planets at the observed gaps in some disks \citep[][]{Pinte2018,Teague2018,Pinte2019,Perez2019b,Pinte2020}, it is still unknown whether such planets also exist in the other disks. If planets exist in the younger disks, planet formation within $\sim 1$ Myr is necessary \citep[e.g.,][]{Sheehan2018}. However, such a fast planet formation at wide orbits ($r\gtrsim 5$ au) still seems difficult at least in the core accretion model \citep[][]{Mizuno1980,Pollack1996}. For example, forming planetary cores necessary to trigger gas accretion onto them takes too long time at radii $\gtrsim10$ au because of collisional fragmentation of planetesimals \citep[][]{Kobayashi2010,Kobayashi2011} even in an initially 10 times more massive disk than the minimum mass solar nebula \citep[][]{Hayashi1981}. Recently, \citet[][]{Ndugu2019} performed the pebble-based planet population synthesis including the type-I and type-II migrations \citep[e.g.,][]{Bitsch2017,Brugger2018,Chambers2018} and examined whether it is possible for all of the gaps observed in Disk Substructures at High Angular Resolution Project \citep[DSHARP,][]{Andrews2018} to be created by unseen growing planets. They showed that large amounts ($\sim 2000\merth$) of pebbles are necessary to reproduce all of the gap structures. It seems quite difficult to form such amount of pebbles in all of the DSHARP disks. Thus, to investigate other ring-gap formation mechanisms is as important as the studies on the planet-induced-ring-gap formation.

Various theories without assuming planets have been proposed to explain the ring-gap formation. \citet[][]{Dullemond2018} investigated the possibility that dust grains are trapped at hypothetical pressure bumps \citep[e.g.,][]{Whipple1972} for the DSHARP objects. The boundary of magnetically dead and active zones is one possible site to set such a pressure bump \citep[][]{Flock2015}. Magnetic activities of a disk-wind system are also found to form ring-like substructures through reconnection of the toroidal magnetic fields \citep[][]{Suriano2018,Suriano2019}. Based on local shearing-box simulations and linear analyses, \citet[][]{Riols2019} shows that a disk-wind system is indeed unstable and host rings and gaps as a result of such a wind-driven instability.

While in the above processes dust grains are passive and follow the background gas structures (e.g., pressure bumps), dust itself can drive processes that lead to the dusty ring formation. Those ``active" processes include secular gravitational instability (GI) \citep[][]{Takahashi2014,Takahashi2016,Tominaga2018}, which is explained in detail below, and two-component viscous GI \citep[TVGI,][]{Tominaga2019}. Both of them require the self-gravity and the dust-gas friction, which causes the angular momentum transport between the dust and gas disks. In contrast to the secular GI, TVGI also requires turbulent gas viscosity that redistributes the gas angular momentum. The presence of dust leads to another type of instability called viscous ring instability \citep[][]{DullemondPenzlin2018}. Instead of the self-gravity, this instability requires the turbulent viscosity that is reduced with increasing dust-to-gas mass ratio \citep[e.g.,][]{Sano2000,Ilgner2006}.

Radial variation of the dust size can also result in the ring-gap structures. \citet[][]{Zhang2015} discusses the possibility that rapid pebble growth near snow lines explains the three prominent gaps in the HL Tau disk. Sintering of dust aggregates is another important process that changes those sizes near the snow lines. \citet[][]{Okuzumi2016} shows that dust aggregates that experience sintering fragment and pile up slightly outside snow lines, and those dust-piling-up regions are observed as bright rings. Those radially changing dust size results in variations of the ionization degree and mass accretion rate across the snow lines, which augments the ring-gap formation \citep[][]{Hu2019}. This process is similar to the instability discussed in \citet[][]{DullemondPenzlin2018}.

This work focuses on the secular GI. As mentioned above, the secular GI can be the origin of the observed multiple rings. Moreover, it is also one of the promising mechanisms for creating planetesimals. Since dust rings formed by the secular GI can be dense and massive, planetesimal formation in the rings via the gravitational instability or coagulation will be expected. Thus, investigating the secular GI gives important clues to reveal the origin of the observed rings and planetesimal formation.

Original linear stability analyses of the secular GI were based on one-fluid equations of a dust disk that is self-gravitating and frictionally interacts with a background Keplerian gas disk \citep[][]{Ward2000,Youdin2005a,Youdin2005b}. The secular GI is triggered by the dust-gas friction as explained below. The dust-gas friction makes the dust follow the background Keplerian flow and prevents the epicyclic motion. In other words, the angular momentum transport between gas and dust reduces the rotational support of the self-gravitating dust disk. This process unconditionally causes the gravitational concentration of the dust and makes the dust disk unstable. Although the secular GI requires the self-gravity, the instability develops even in self-gravitationally stable dust disks. As discussed in the previous studies \citep[][]{Youdin2005a,Youdin2011,Tominaga2019}, the secular GI is distinct from the frictionally mediated GI and originates from a static mode\footnote{This kind of mode is referred to as a neutral mode in \citet[][]{Youdin2005a,Youdin2011}} that is a steady solution of the perturbation equations in the absence of the friction \citep[see also Figure 1 of ][]{Tominaga2019}.

Previous studies have shown that the dust diffusion due to gas turbulence stabilizes the secular GI. \citet[][]{Youdin2011}, \citet[][]{Shariff2011} and \citet[][]{Michikoshi2012} analyzed the linear stability of the secular GI in the presence of the dust diffusion. Although the diffusion suppresses the instability at short wavelengths, \citet[][]{Youdin2011} concluded that dust rings containing up to $\sim 0.1\;\merth$ collapse and would fragment into planetesimals. The analysis by \citet[][]{Michikoshi2012} is based on the stochastic diffusion model, which is more rigorous than the hydrodynamic models used in \citet[][]{Youdin2011} and \citet[][]{Shariff2011}, and they obtained the essentially same results. 

\citet[][]{Takahashi2014} took into account the back reaction from dust to gas and performed linear analyses. The back reaction is found to stabilize long-wavelength perturbations and renders the secular GI operational at intermediate wavelengths. They also suggested that the secular GI grows and creates rings even at a radius of 100 au in weakly turbulent disks in which the so-called $\alpha$ value \citep[][]{Shakura1973} is roughly less than $10^{-3}$ or $10^{-4}$. This also implies that the secular GI is a promising mechanism to form planetesimals composing a debris disk. The growth timescale of the ``two-component" secular GI is longer than the Keplerian period by orders of magnitude \citep[see also][]{Latter2017}, which could be shorter if multiple dust species or magnetic fields are considered \citep[][]{Shadmehri2016a,Shadmehri2016b}. 

The above studies used the advection-diffusion equation for dust surface density to model the dust diffusion. It is known that such a modeling has the momentum conservation break down \citep[e.g.,][]{Goodman2000}. \citet[][]{Tominaga2019} revised the previous equations in order to recover the conservation law and reanalyzed the secular GI. While the previous studies found the secular GI overstable \citep[e.g.,][]{Youdin2011,Takahashi2014,Latter2017}, the reanalyses by \citet[][]{Tominaga2019} showed that the secular GI is actually a monotonically growing mode, and the overstabilization found in the previous work is due to the unphysical change of the angular momentum due to the diffusion modeling \citep[see Figure 2 of][]{Tominaga2019}. Although the mode properties of the secular GI is revised, \citet[][]{Tominaga2019} suggested that the instability still has the potential to form planetesimals and multiple rings/gaps in a protoplanetary disk.

Nonlinear growth of the secular GI was investigated for the first time in our previous study \citep[][]{Tominaga2018}. We performed numerical simulations and found that in the nonlinear growth stage dust rings collapse self-gravitationally while gas avoid the collapse because of the pressure gradient force, which results in high dust-to-gas mass ratio in the resultant dust rings (see Figure 8 therein). However, we did not include two physical processes in our simulations: the dust diffusion and the radial drift of the dust. Since the dust diffusion is the most efficient process to stabilize the secular GI \citep[][]{Youdin2011}, it will also affect the nonlinear growth and its saturation. If the nonlinear growth is saturated by the dust diffusion, the dust-to-gas mass ratio in the resultant dust rings will be smaller than our previous study showed. The radial drift of the dust is also important process. It is still unclear whether the secular GI can grow in the presence of the drift motion between dust and gas although ``one-component'' secular GI will grow as mentioned by \citet{Youdin2005a}. 

In this work, we performed numerical simulations of the secular GI in a radially extended dusty gas disk with taking into account the dust diffusion and the radial drift. We investigate effects of the above two processes on the nonlinear growth of the secular GI, especially the dust-to-gas mass ratio in the resultant rings.

This paper is organized as follows. In Section \ref{sec:method}, we summarize basic equations describing the macroscopic dynamics, our numerical method and setups. We overview the results of simulations in Section \ref{sec:result}, and give its physical interpretation based on the linear analyses and implication to the multiple ring formation in a global disk in Section \ref{sec:discussion}. Section \ref{sec:conclusion} presents conclusions with a brief summary.

\section{Method}\label{sec:method}
\subsection{Basic Equations}\label{sec:basiceq}
We consider time evolution of axisymmetric dust and gas disks. Since the secular GI requires the dust-gas friction, the dust layer around the disk midplane seems to be the most important region. The gas above the dust layer will also contribute to the growth of the secular GI through the gravitational interaction while its frictional interaction with the dust is relatively weak. Thus, we consider a ``lower layer" that includes those dust and gas driving the secular GI. Hereafter, we do not concern the vertical extent of the lower layer since it is beyond the scope of this study. The following equations govern the evolution of physical quantities vertically integrated within the lower layer orbiting around a central star with mass $M_{\ast}$ in the cylindrical coordinate $(r,\phi)$ : 
\begin{equation}
\frac{\partial \Sigma}{\partial t}+\frac{1}{r}\frac{\partial}{\partial r}\left(r\Sigma u_r\right)=0,\label{eq:eocgas}
\end{equation}
\begin{align}
\Sigma\left(\frac{\partial u_r}{\partial t}+u_r\frac{\partial u_r}{\partial r}\right)=&-\frac{\partial}{\partial r}\left(\cs^2\Sigma\right)-\Sigma\frac{\partial}{\partial r}\left(\Phi-\frac{GM_{\ast}}{r}\right)\notag\\
&+\frac{2}{r}\frac{\partial}{\partial r}\left(r\Sigma\nu\frac{\partial u_r}{\partial r}\right)-\frac{2\Sigma\nu u_r}{r^2}\notag\\
&-\frac{\partial}{\partial r}\left(\frac{2\Sigma\nu}{3r}\frac{\partial ru_r}{\partial r}\right)\notag\\
&+\sigmad\frac{v_r-u_r}{\tstop},\label{eq:eomrgas}
\end{align}
\begin{align}
\Sigma\left(\frac{\partial j_{\gas}}{\partial t}+u_r\frac{\partial j_{\gas}}{\partial r}\right)=&\frac{1}{r}\frac{\partial}{\partial r}\left[r^3\Sigma\nu\frac{\partial}{\partial r}\left(\frac{u_{\phi}}{r}\right)\right]\notag\\
&+\sigmad\frac{j_{\dst}-j_{\gas}}{\tstop},\label{eq:eomphigas}
\end{align}
\begin{equation}
\frac{\partial\sigmad}{\partial t}+\frac{1}{r}\frac{\partial}{\partial r}\left( r\sigmad v_r\right)=\frac{1}{r}\frac{\partial}{\partial r}\left(rD\frac{\partial\sigmad}{\partial r}\right),\label{eq:eocdust}
\end{equation}
\begin{align}
\sigmad\biggl[\frac{\partial v_r}{\partial t}+&\left(v_r-\frac{D}{\sigmad}\frac{\partial\sigmad}{\partial r}\right)\frac{\partial v_r}{\partial r}\biggr]\notag\\
=&\sigmad\frac{v_{\phi}^2}{r}-\frac{\partial}{\partial r}\left(\cd^2\sigmad\right)-\sigmad\frac{\partial}{\partial r}\left(\Phi-\frac{GM_{\ast}}{r}\right)\notag\\
&-\sigmad\frac{v_r-u_r}{\tstop}+\frac{1}{r}\frac{\partial}{\partial r}\left(rv_rD\frac{\partial\sigmad}{\partial r}\right),\label{eq:eomrdust}
\end{align}
\begin{equation}
\sigmad\left[\frac{\partial j_{\dst}}{\partial t}+\left(v_r-\frac{D}{\sigmad}\frac{\partial\sigmad}{\partial r}\right)\frac{\partial j_{\dst}}{\partial r}\right]=-\sigmad\frac{j_{\dst}-j_{\mathrm{g}}}{\tstop},\label{eq:eomphidust}
\end{equation}
\begin{equation}
\nabla^2\Phi=4\pi G\left(\Sigma+\sigmad\right)\delta(z),\label{eq:poisson}
\end{equation}
where $\Sigma$ and $\sigmad$ are the gas and dust surface densities of the lower layer, $u_i$ and $v_i$ are the $i$th component of the velocities, $j_{\mathrm{g}}$ and $j_{\dst}$ denote the specific angular momentums of the gas and the dust, $\cs$ is the sound speed, $\cd$ and $D$ are the velocity dispersion and the diffusion coefficient of the dust, respectively. The self-gravitational potential is denoted by $\Phi$, and the gravitational constant is $G$. We denote the coefficient of the turbulent viscosity by $\nu$ measured by the dimensionless parameter $\alpha\equiv\nu\Omega\cs^{-2}$ \citep{Shakura1973}, where $\Omega$ is the angular velocity of the gas disk.

The terms including $D\partial\sigmad/\partial r$ in Equations (\ref{eq:eomrdust}) and (\ref{eq:eomphidust}) are introduced by \citet[][]{Tominaga2019}. Those terms are derived from the mean-field approximation and represent the mean momentum flow caused by the diffusion. By including those terms, both total linear and angular momentums are conserved \citep[see also Appendix A of][]{Tominaga2019}. The last term on the right hand side of Equation (\ref{eq:eomrdust}) affects the dynamics if $\alpha\gtrsim 10^{-3}$.

\subsection{Numerical Methods and Setups}\label{sec:numerics}
Since the growth timescale of the secular GI is much longer than the Keplerian periods, numerical errors due to the time integration need to be minimized as possible. In addition to this, we should also minimize numerical viscosity that unphysically smooths out advecting density fluctuations. As we will show in Section \ref{sec:discussion}, the secular GI originates from the ``dust-streaming mode" where the dust surface density perturbation propagates with the radial drift velocity. Thus, we need to treat the advection of dust with keeping numerical dissipation minimal as much as possible. 

For those reasons, we use a numerical method that incorporates the symplectic integrator and the Lagrangian-cell method \citep[][]{Tominaga2018}. The symplectic integrator enables us to avoid a monotonic increase of the error due to the time integration. The Lagrangian-cell method makes it possible to properly deal with the advection without numerical dissipation due to the spatial discretization. In this scheme, density and pressure are defined at cell centers, and velocity and angular momentum are defined at cell boundaries. We use $m_{\dst,i}$, $r_{\dst,i}$, $m_{\gas,i}$ and $r_{\gas,i}$ to denote masses and radial positions of the $i$th dust and gas cells (rings), respectively. Mass $m_{s,i+1/2}\equiv (m_{s,i+1}+m_{s,i})/2$ is also assigned at a cell boundary at $r_s=r_{s,i+1/2}$ between $i$th and $(i+1)$th cells, where $s=\dst, \gas$. The gas and dust velocities and angular momentums are defined at the cell boundaries and denoted by $u_{r,i+1/2},\;v_{r,i+1/2},\;j_{\gas,i+1/2}$ and $j_{\dst,i+1/2}$. We note that the position of a dust cell does not necessarily coincide with that of a gas cell.

We adopted the operator-splitting technique with the second-order accuracy in time \citep[][]{Inoue2008}. The basic equations are split into four parts: (A) the frictional interaction part, (B) the dust diffusion part, which includes the last term on the right hand side of Equation (\ref{eq:eomrdust}), (C) the gas viscosity part, and (D) the hydrodynamic part with the other physical processes for gas and dust. 

The parts (A) and (D) follows \citet{Tominaga2018}. In the part (A), the frictional interaction between dust and gas is solved with the piecewise exact solution, which is free from the limitation on the time step $\Delta t$ due to the small stopping time $\tstop$. In this step, we use interpolation functions shown in Appendix 3 of \citet{Tominaga2018} for the radial velocities and the specific angular momentums to calculate the frictional interaction. In the part (D), we calculate the forces (e.g., pressure gradient force, self-gravity) acting on the dust and gas cell boundaries and updates their positions and velocities with the leap-frog integrator. The self-gravity is represented as a superposition of infinitesimally thin ring gravity \citep[see,][]{Tominaga2018}. We weakens the ring gravity with a softening length of half cell's width ($\sim0.1$ au), and rescale it when the cell width becomes larger in time. 

We newly implement the parts (B) and (C) based on the second-order Runge-Kutta integrator. In the part (B), we first interpolate the dust surface density using a quadratic function to evaluate the mass flux $D\partial\sigmad/\partial r$ at the cell boundaries. We then displace the dust cell boundaries using velocity $-\sigmad^{-1}D\partial\sigmad/\partial r$, which corresponds to the diffusion term in Equation (\ref{eq:eocdust}). We do not change the angular momentums of the dust cell boundaries when displacing the dust since the Lagrange derivative in the equation of motion includes the advection along the diffusion flow (Equation (\ref{eq:eomphidust})). On the other hand, the radial linear momentum of the $i$th cell boundary, $m_{\dst,i+1/2}v_{r,i+1/2}$, changes according to the last term on the right hand side of Equation (\ref{eq:eomrdust}). We evaluate $F(r)\equiv rv_rD\partial\sigmad/\partial r$ at $r=r_{\dst,i+1},\;r_{\dst,i}$, and update $m_{\dst,i+1/2}v_{r,i+1/2}$ with using $F(r_{\dst,i+1})-F(r_{\dst,i})$. 

In the part (C), we interpolate the radial velocity $u_r$ and the specific angular momentum $j_{\gas}$ in order to obtain the radial and angular momentum fluxes due to the gas viscosity at $r=r_{\gas,i+1},\;r_{\gas,i}$. From Equations (\ref{eq:eomrgas}) and (\ref{eq:eomphigas}), the radial and angular momentum changes are given by 
\begin{equation}
\left(\frac{\partial\Sigma u_r}{\partial t}\right)_{\mathrm{vis}}\equiv\frac{1}{r}\frac{\partial F_{\mathrm{vis},r}}{\partial r}-\frac{2u_r}{3r^3}\frac{\partial r^2\Sigma\nu}{\partial r},
\end{equation}
\begin{equation}
\left(\frac{\partial\Sigma j_{\gas}}{\partial t}\right)_{\mathrm{vis}}\equiv\frac{1}{r}\frac{\partial F_{\mathrm{vis},\phi}}{\partial r},
\end{equation}
\begin{equation}
F_{\mathrm{vis},r}\equiv\frac{4}{3}r\Sigma\nu\frac{\partial u_r}{\partial r},
\end{equation}
\begin{equation}
F_{\mathrm{vis},\phi}\equiv r^3\Sigma\nu\frac{\partial}{\partial r}\left(\frac{u_{\phi}}{r}\right).
\end{equation}
Integrating the above equations in the radial and azimuthal directions, we obtain the following equations for updating $m_{\gas,i+1/2}u_{r,i+1/2}$ and $m_{\gas,i+1/2}j_{\gas,i+1/2}$:
\begin{align}
\frac{\partial m_{\gas,i+1/2}u_{r,i+1/2}}{\partial t}=&\left[2\pi F_{\mathrm{vis},r}\right]^{r_{\gas,i+1}}_{r_{\gas,i}}\notag\\
&-\frac{4\pi}{3}\int^{r_{\gas,i+1}}_{r_{\gas,i}}\frac{u_r}{r^2}\frac{\partial r^2\Sigma\nu}{\partial r}dr,\label{eq:dmurvis}
\end{align}
\begin{equation}
\frac{\partial m_{\gas,i+1/2}j_{\gas,i+1/2}}{\partial t}=\left[2\pi F_{\mathrm{vis},\phi}\right]^{r_{\gas,i+1}}_{r_{\gas,i}},
\end{equation}
where $[A(r)]^{r_2}_{r_1}\equiv A(r_2)-A(r_1)$. We evaluate the last term on the right hand side of Equation (\ref{eq:dmurvis}) by $-\left(4\pi u_{r,i+1/2}/3r_{\gas,i+1/2}^2\right)\left[r^2\Sigma\nu\right]^{r_{\gas,i+1}}_{r_{\gas,i}}$.

We determine the time step based on the following:
\begin{equation}
\Delta t=\mathrm{min}\left(\Delta t_{\gas},\Delta t_{\dst},\Delta t_{\mathrm{diff}},\Delta t_{\mathrm{vis}}\right),
\end{equation}
where $\Delta t_{\gas}$ and $\Delta t_{\dst}$ are the time steps determined by the CFL condition for the gas and the dust equations without the dust diffusion or the viscosity, $\Delta t_{\mathrm{diff}}$ and $\Delta t_{\mathrm{vis}}$ are those determined by the diffusion term and the viscosity term: $\Delta t_{\mathrm{diff}}=0.125\times\mathrm{min}((r_{\dst,i+1/2}-r_{\dst,i-1/2})^2/D)$, and $\Delta t_{\mathrm{vis}}=0.125\times\mathrm{min}((r_{\gas,i+1/2}-r_{\gas,i-1/2})^2/\nu)$. As shown in the next section, the dust rings become spiky as the secular GI grows, and then the dust diffusion tends to limit $\Delta t$. Thus, we adopt the super-time-stepping (STS) \citep[][]{Alexiades1996} in the part (B) when the time step $\Delta t$ is limited by the dust diffusion. The number of substeps and the stability parameter in STS are fixed to be 4 and 0.1, respectively. We just moderately accelerate the time stepping since the physical diffusion timescale becomes short ($\ll$ one Keplerian period) and the total time step in STS should not exceed it.

The number of cells $N_r$ is 1024 for both dust and gas. The initial position of the inner boundaries are 10 au. We space the dust and gas domains so that each cell has the same mass and the outer boundaries are located at $r\simeq 300$ au. We fixed the gas and dust outer boundaries in simulations. The gas inner boundary moves so that the inner most gas surface density is constant in time. We have dust cells at $r<r_{\gas,1}$ move inward with the steady drift velocity \citep[e.g.,][]{Nakagawa1986} estimated with the initial density profiles at $r=r_{\gas,1}$.

\subsection{Disk models and parameters}\label{sec:setups}
We consider gas and dust disks around $1\msun$ mass star, and those initial surface density profiles is given by the following power law function:
\begin{equation}
\Sigma(r)=\Sigma_{100}\left(\frac{r}{100\;\mathrm{au}}\right)^{-q}\exp\left(-\frac{r}{100\;\mathrm{au}}\right),\label{eq:gasdiskmodel}
\end{equation}
\begin{equation}
\sigmad(r)=\Sigma_{\dst,100}\left(\frac{r}{100\;\mathrm{au}}\right)^{-q}\exp\left(-\frac{r}{100\;\mathrm{au}}\right),
\end{equation}
where $\Sigma_{100}$ and $\Sigma_{\dst,100}$ are constants. 
In this study, we use the Toomre's $Q$ value of the gas, $Q=\cs\Omega/\pi G\Sigma$, to show how massive the lower layer is. One obtains $\Sigma_{100}$ from the $Q$ value at $r=100\;\mathrm{au}$ (Table \ref{tab:param}), Keplerian angular velocity and the temperature shown below. The dust surface density at $r=100$ au is determined by the assumed initial dust-to-gas ratio in the lower layer $\sigmad/\Sigma$. We note that $\Sigma_{\dst}/\Sigma$ represents the dust-to-gas mass ratio averaged in the lower layer, which is different from $\Sigma_{\dst,\mathrm{tot}}/\Sigma_{\mathrm{tot}}$ where $\Sigma_{\dst,\mathrm{tot}}$ and $\Sigma_{\mathrm{tot}}$ are total surface densities of the dust and gas disks including both upper and lower layers. In weakly turbulent gas disks, $\sigmad/\Sigma$ easily becomes higher than $\Sigma_{\dst,\mathrm{tot}}/\Sigma_{\mathrm{tot}}$ by an order of magnitude. 
In this work, considering dust rich disks $\Sigma_{\dst,\mathrm{tot}}/\Sigma_{\mathrm{tot}}=0.05$, we assume $\sigmad/\Sigma=0.1$. 
The power-law index is one of the important parameters that characterize the dust drift. \citet{Kitamura2002} observed 13 disks around T Tauri stars and found that the power-law index is 0-1 in most cases. Motivated by this study, we take the median value and fix $q=0.5$ in this work. 

The gas disk in our simulations is locally isothermal, and the temperature profile $T(r)$ is
\begin{equation}
T(r)=10\;\mathrm{K}\left(\frac{r}{100\;\mathrm{au}}\right)^{-1/2},\label{eq:tempmodel}
\end{equation}
where we mimic a disk passively heated by the stellar radiation \citep[e.g.,][]{Chiang1997}. We calculate the sound speed $\cs$ at each radius assuming the molecular weight to be 2.34. The gas scale height $H\equiv\cs/\Omega$ is $H\simeq 6.3\;\mathrm{au}\left(r/100\mathrm{au}\right)^{5/4}$. The initial azimuthal velocity is determined by the radial force balance without using the friction, the diffusion or the viscosity. We put initially random perturbations to the cell positions and the velocities. Amplitudes of the position and velocity perturbations are five percent of the cell widths and $\cd$, respectively.

The normalized stopping time $\tstop\Omega$ is an important parameter characterizing the growth rate and the unstable wavelengths of the secular GI. A timescale of the dust coagulation is shorter than that of the secular GI when dust grains are small ($\sim 100$ orbits), and thus dust grains would grow up to the drift-limited size \citep[see also,][]{Okuzumi2012}.  We mimic this situation by simply setting constant $\tstop\Omega$. Assuming the total dust-gas mass ratio $\Sigma_{\dst,\mathrm{tot}}/\Sigma_{\mathrm{tot}}$, one can estimate the drift-limited stopping time. In Appendix \ref{ap:tstop}, we derive the drift-limited stopping time based on our disk model and show that for $\Sigma_{\dst,\mathrm{tot}}/\Sigma_{\mathrm{tot}}=0.05$ the drift-limited $\tstop\Omega$ is about 0.6. 
We do not consider fragmentation-limited dust sizes because we are focusing on dynamics at outer region in weakly turbulent gas disks \citep[see also,][]{Birnstiel2009,Birnstiel2012}. 

Although the diffusion coefficient $D$ and the velocity dispersion $\cd$ depend on $\tstop\Omega$ and $\alpha$ \citep[][]{YL2007}, those are well approximated by $D\simeq\alpha\cs^2\Omega^{-1}$ and $\cd\simeq\sqrt{\alpha}\cs$ for $\tstop\Omega<1$. Thus we use the relations $D=\alpha\cs^2\Omega^{-1}$ and $\cd=\sqrt{\alpha}\cs$ in the present simulations for simplicity. In this study, we perform numerical simulations with different Toomre's $Q$ values and turbulent strength $\alpha$. We summarize the parameters in Table \ref{tab:param}. 

Our choice of parameters for the density distribution is optimistic to study the global and nonlinear outcomes of the secular GI. For example, disks are massive (see also Table \ref{tab:massfrac}). On the other hand, the disk masses and the dust-to-gas surface density ratio $\Sigma_{\dst,\mathrm{tot}}/\Sigma_{\mathrm{tot}}$ are not observationally well-constrained because of uncertainty of the dust opacity. In addition, recent work shows that neglecting the scattering effect of the dust thermal emission underestimates the dust mass in a disk \citep[][]{Zhu2019}. In the present study, we thus assume massive disks where the secular GI relatively easily grows.

\section{Results}\label{sec:result}
In this section, we overview the results of our simulations. We identify two regimes in the disk evolution via the secular GI: formation of ``thin dense dust rings" and ``transient low-contrast dust rings". The results are briefly summarized in Table \ref{tab:param}. The following subsections show those results in detail.

\begin{deluxetable*}{c|cccc}[ht]
\tablecaption{Summary of parameters and results\label{tab:param}}
\tablehead{
\colhead{Label} & \colhead{$Q_{100}$\tablenotemark{a}} & \colhead{$\alpha$\tablenotemark{b}} & \colhead{Results\tablenotemark{c}} & \colhead{Time that simulations last for} 
}
\startdata
Q4a10 & 4 & $1\times 10^{-3}$ & thin dense dust rings & $2.1 \times 10^{4}$ yr\\
Q4a20 & 4 & $2\times 10^{-3}$ & transient low-contrast rings & $5.9 \times 10^{4}$ yr\\
Q5a5 & 5 & $5\times 10^{-4}$ & thin dense dust rings & $1.9 \times 10^{4}$ yr\\
Q5a8 & 5 & $8\times 10^{-4}$ & transient low-contrast dust rings & $5.6 \times 10^{4}$ yr\\
Q5a8L\tablenotemark{d} & 5 & $8\times 10^{-4}$ & thin dense dust rings & $2.5 \times 10^{4}$ yr\\
Q6a3 & 6 & $3\times 10^{-4}$ & thin dense dust rings & $2.7 \times 10^{4}$ yr\\
Q6a5 & 6 & $5\times 10^{-4}$ & transient low-contrast dust rings  & $6.2 \times 10^{4}$ yr\\
\enddata
\tablenotetext{a}{Toomre's $Q$ value for the lower layer of a gas disk at $r=$100 au}
\tablenotetext{b}{The strength of turbulence}
\tablenotetext{c}{Results of simulations: formation of ``thin  dense dust rings" or ``transient low-contrast dust rings"}
\tablenotetext{d}{The letter ``L" means a run with six times larger perturbations (see Section \ref{subsec:condition})}
\end{deluxetable*}
\begin{figure*}
	\begin{tabular}{c}
		\begin{minipage}{0.5\hsize}
			\begin{center}
				\includegraphics[width=1.0\columnwidth]{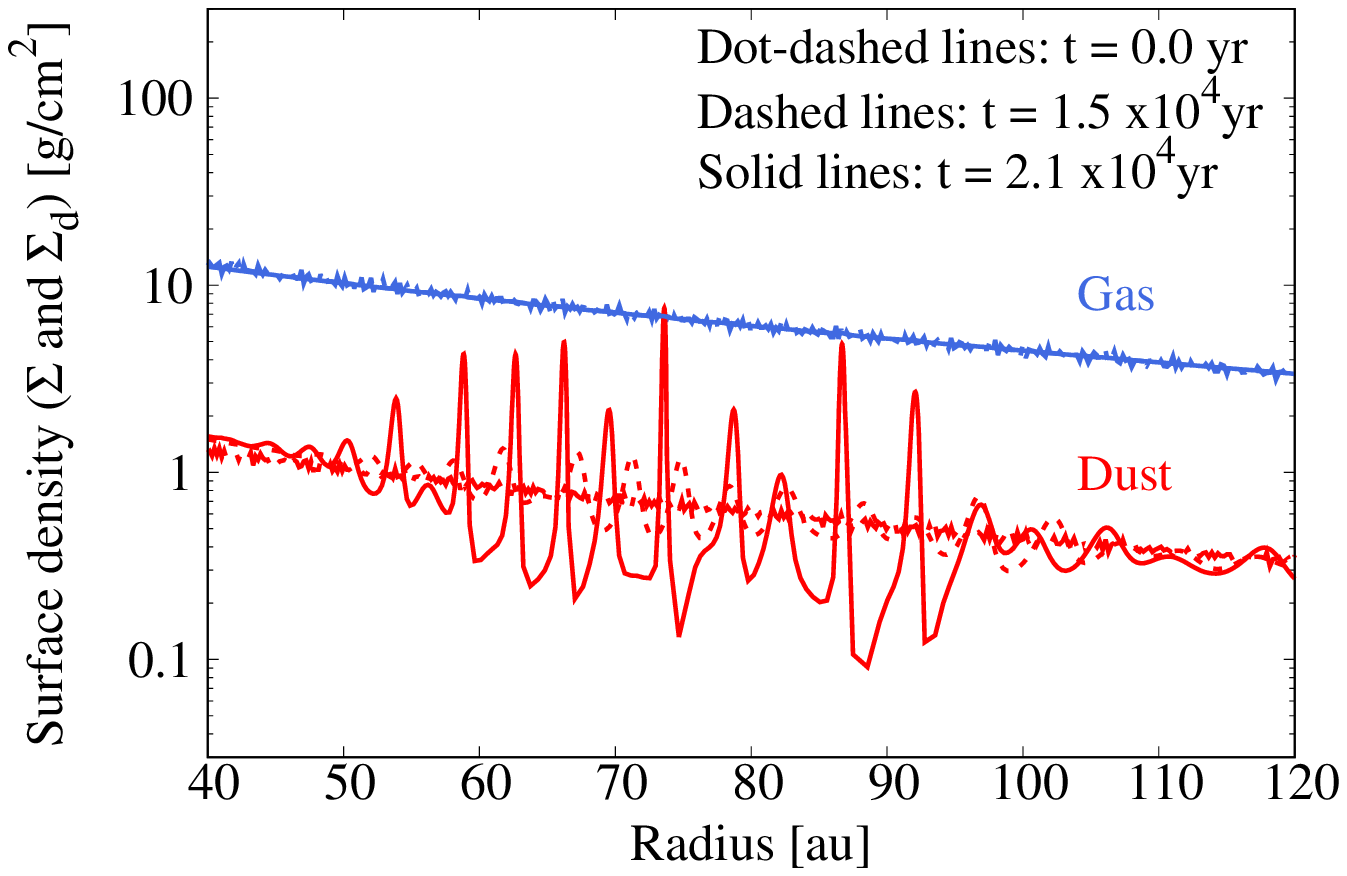}
			\end{center}
		\end{minipage}
		\begin{minipage}{0.5\hsize}
			\begin{center}
				\includegraphics[width=1\columnwidth]{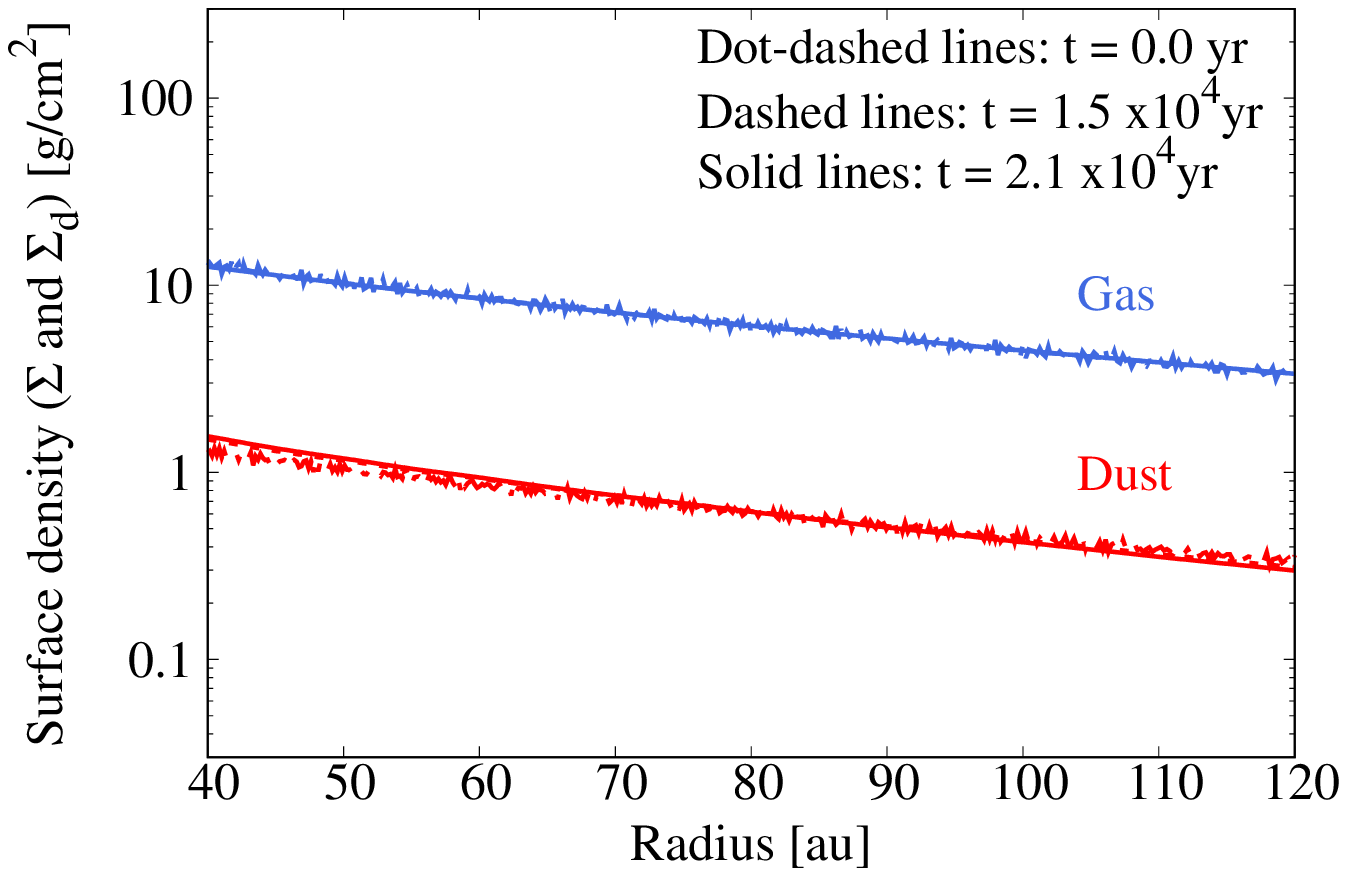}
			\end{center}
		\end{minipage}
	\end{tabular}
\caption{(Left panel) Surface density evolution from Q4a10 run. Multiple rings and gaps form within $10^4$ yr. (Right panel) Surface density evolution from a run where we use the same parameters as in Q4a10 run and switch off the self-gravity of the dust and gas disks. In both panels, red and blue lines show the dust and gas surface densities. The dotted-dashed, dashed and solid lines show snapshots at $t=$0.0 yr, $1.5\times10^4$ yr and $2.1\times10^4$ yr, respectively. We note that the classical GI is stabilized by the dust diffusion, and thus the multiple ring-gap formation is due to the secular GI.}
\label{fig:sigmaevolv_Q4a10ep01}
\end{figure*}
\begin{figure*}
	\begin{tabular}{c}
		\begin{minipage}{0.5\hsize}
			\begin{center}
				\includegraphics[width=1.0\columnwidth]{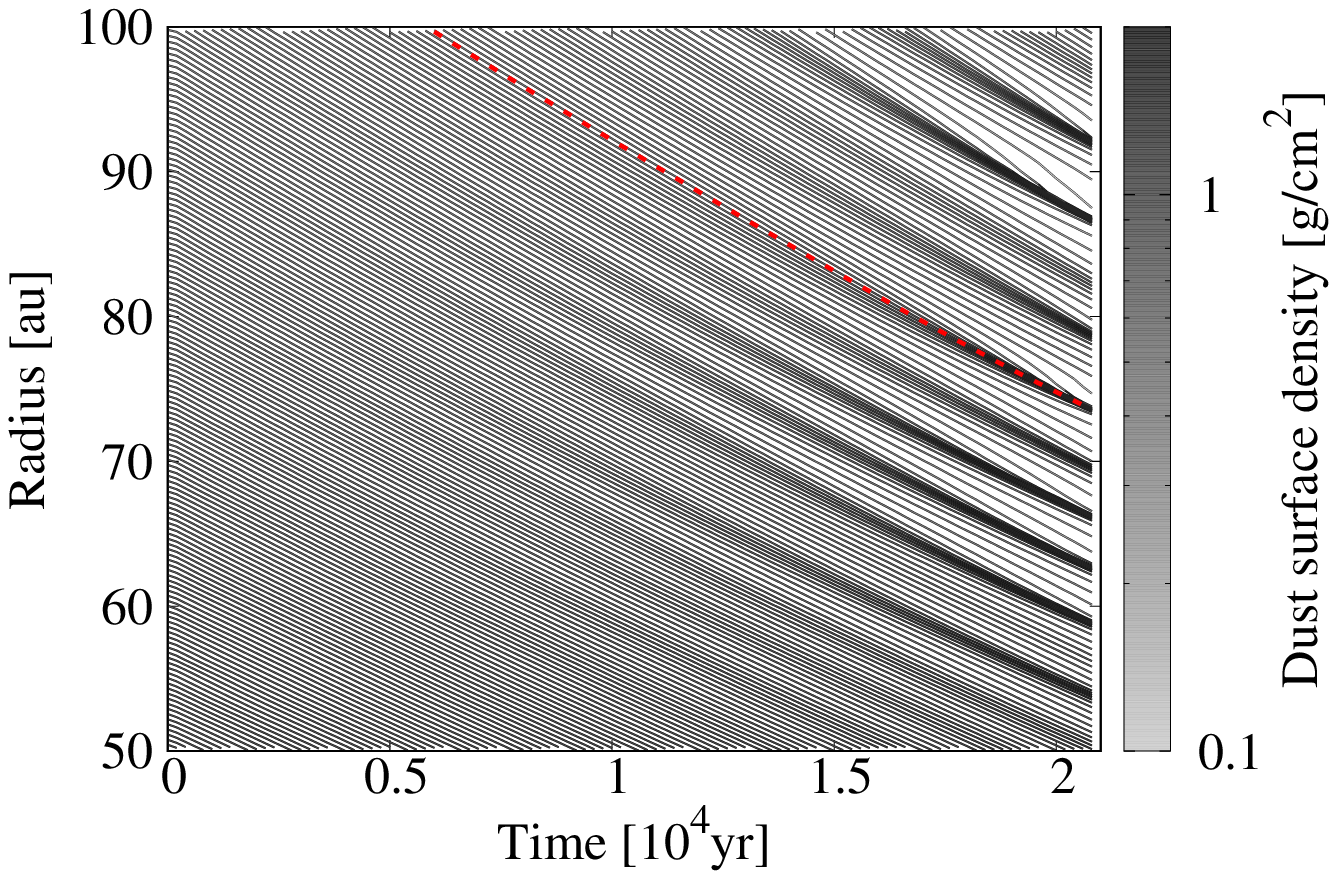}
			\end{center}
		\end{minipage}
		\begin{minipage}{0.5\hsize}
			\begin{center}
				\includegraphics[width=1\columnwidth]{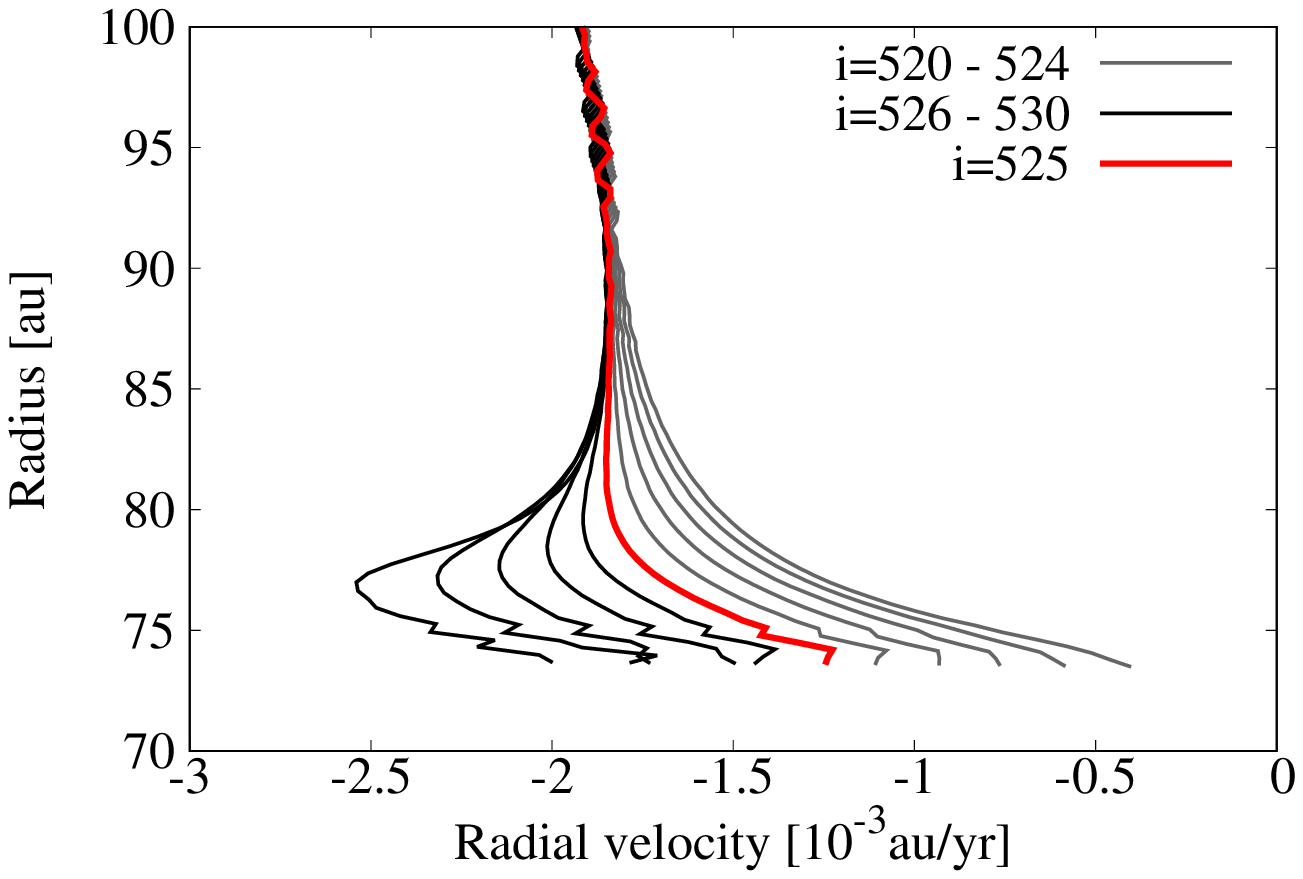}
			\end{center}
		\end{minipage}
	\end{tabular}
\caption{(Left) Time evolution of the dust-cell positions in Q4a10 run. Color of the lines shows the dust surface density at each dust cell. We used a reduced number of cells in plotting this figure. Red dashed line is the motion of the 525th dust cell. (Right) The radial velocity of some dust cell boundaries ($r=r_{\dst,i+1/2}$ where $i=520-530$) at each radius. The 525th cell boundary shown in red roughly corresponds to the ring peak position. The inner and outer cells are shown in gray and black. The spreading trend in $v_r-r$ plane represents the collapsing motion toward the ring center. At the radius $\simeq 77$ au, the gravitational softening term suppresses the accelerated collapse. At $r\lesssim77$ au, the drift speed decreases as $\sigmad/\Sigma$ increases.} 
   		 \label{fig:t-r-sigma_Q4a10ep01}
\end{figure*}
\begin{figure}[htp]
	\begin{center}
		\includegraphics[width=1.0\columnwidth]{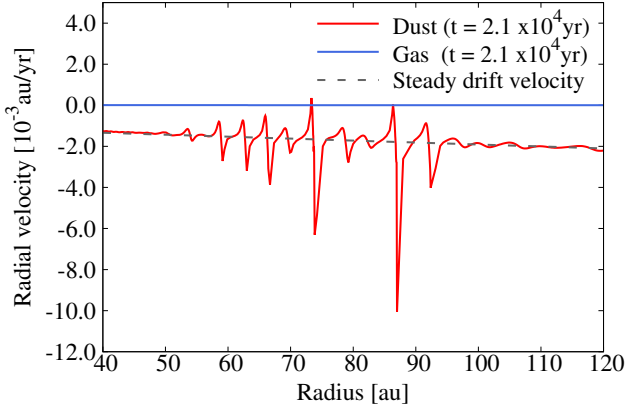}
	\end{center}
    		\caption{Radial velocity profile at $t=2.1\times10^4$ yr in Q4a10 run. Red and blue lines show the dust and gas radial velocities, respectively. The gray dashed line shows the steady drift velocity of dust $v_{\mathrm{dri}}$ \citep[][]{Nakagawa1986}. The mean radial velocity of the dust is in good agreement with the steady drift velocity. The gas has small positive velocity because of the frictional back-reaction on the dust drift.} 
   		 \label{fig:vrQ3a5ep01}
\end{figure}
\begin{figure}[htp]
	\begin{center}
		\includegraphics[width=1.0\columnwidth]{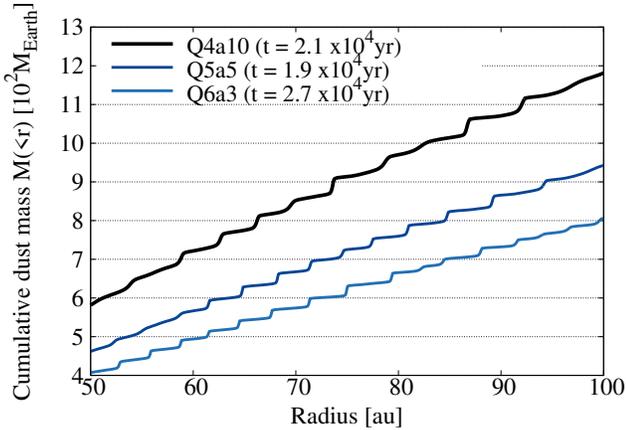}
	\end{center}
    		\caption{Cumulative dust mass $M(<r)$ obtained from Q4a10, Q5a5 and Q6a3 runs. Thick black line corresponds to Q4a10 run, from which we obtain the surface density evolution in Figure \ref{fig:sigmaevolv_Q4a10ep01}. Positions of cliffs correspond to the dust ring radii, and the height of the cliff corresponds to the ring mass. The mass of the individual dust ring is $\gtrsim10\merth$.} 
   		 \label{fig:cumumass}
\end{figure}

\begin{figure}[htp]
	\begin{center}
		\includegraphics[width=1.0\columnwidth]{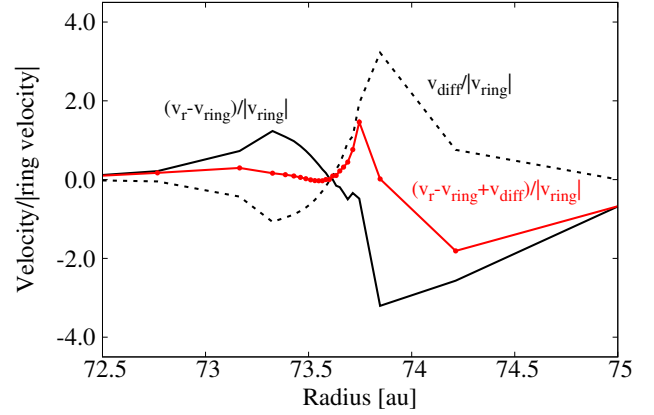}
	\end{center}
    		\caption{Radial velocity profile normalized by the ring velocity $v_{\mathrm{ring}}$ at $t=2.1\times10^4$ yr in Q4a10 run. Black solid line shows the dust velocity with respect to the ring velocity, $v_r-v_{\mathrm{ring}}$. Black dashed line shows the velocity $v_{\mathrm{diff}}\equiv-D\sigmad^{-1}\partial\sigmad/\partial r$. Red line with filled circles shows the sum $v_r-v_{\mathrm{ring}}+v_{\mathrm{diff}}$. The filled circles are the data points.}
   		 \label{fig:vrdiffQ3a5ep01}
\end{figure}

\begin{figure}[htp]
	\begin{center}
		\includegraphics[width=1.0\columnwidth]{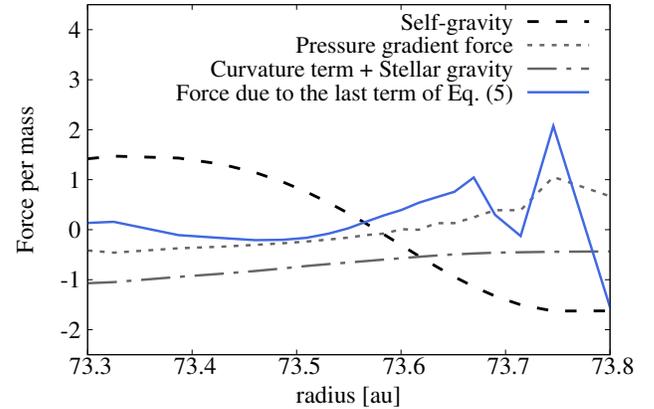}
	\end{center}
    		\caption{Radial profile of the forces per mass acting on the dust in a ring whose radius is $\simeq 73.6$ au. The vertical axis is normalized by $10^{-6}G\msun/(1\;\mathrm{au})^2$. We only plot the self-gravity (black dashed line), the pressure gradient force ($\sigmad^{-1}\cd^2\partial\sigmad/\partial r$, gray short dashed line), the sum of the curvature term and the stellar gravity (gray dot-dashed line), and the force coming from the last term on the right hand side of Equation (\ref{eq:eomrdust}), i.e., $\sigmad^{-1}r^{-1}\partial F(r)/\partial r$ where $F(r)=rv_rD\partial\sigmad/\partial r$ (blue solid line). At the outer half of the ring, the third term becomes comparable to the self-gravity, and decelerates the radial converging speed toward the ring.}
   		 \label{fig:forcecomp}
\end{figure}
\begin{deluxetable*}{c|ccc|ccc}[ht]
\tablecaption{List of the evaluated masses\label{tab:massfrac}}
\tablehead{
 & \multicolumn{3}{c}{$R_{\dst,\mathrm{out}}=120$ au} & \multicolumn{3}{c}{$R_{\dst,\mathrm{out}}=200$ au} 
}
\startdata
Label and Time& $M_{\mathrm{ring,tot}}$ [$M_{\earth}$]& $M_{\dst}$ [$M_{\earth}$]& $M_{\mathrm{ring,tot}}/M_{\dst}$& $M_{\mathrm{ring,tot}}$ [$M_{\earth}$]& $M_{\dst}$ [$M_{\earth}$]& $M_{\mathrm{ring,tot}}/M_{\dst}$\\\hline
Q4a10 ($t=2.0\times 10^4$ yr) & 4.7$\times10^{2}$& 1.2$\times10^{3}$& 0.38 & 8.5$\times10^{2}$& 1.8$\times10^{3}$& 0.46  \\
Q5a5 ($t=1.6\times 10^4$ yr) & 4.5$\times10^{2}$& 9.8$\times10^{2}$& 0.46 & 7.4$\times10^{2}$& 1.4$\times10^{3}$& 0.51 \\
Q5a8L ($t=2.3\times 10^4$ yr) & 3.2$\times10^{2}$& 9.8$\times10^{2}$& 0.33 & 6.9$\times10^{2}$& 1.4$\times10^{3}$& 0.47  \\
Q6a3 ($t=1.8\times 10^4$ yr)& 3.5$\times10^{2}$ & 8.1$\times10^{2}$& 0.42 & 7.3$\times10^{2}$& 1.2$\times10^{3}$& 0.60  \\
\enddata
\end{deluxetable*}

\begin{deluxetable*}{c|ccc|ccc}[ht]
\tablecaption{List of the evaluated masses\label{tab:massfrac_ri60}}
\tablehead{
 & \multicolumn{3}{c}{$R_{\dst,\mathrm{in}}=60\;\mathrm{au},\;R_{\dst,\mathrm{out}}=120\;\mathrm{au}$} & \multicolumn{3}{c}{$R_{\dst,\mathrm{in}}=60\;\mathrm{au},\;R_{\dst,\mathrm{out}}=200\;\mathrm{au}$} 
}
\startdata
Label and Time& $M_{\mathrm{ring,tot}}$ [$M_{\earth}$]& $M_{\dst}$ [$M_{\earth}$]& $M_{\mathrm{ring,tot}}/M_{\dst}$& $M_{\mathrm{ring,tot}}$ [$M_{\earth}$]& $M_{\dst}$ [$M_{\earth}$]& $M_{\mathrm{ring,tot}}/M_{\dst}$\\\hline
Q4a10 ($t=2.0\times 10^4$ yr) & 4.7$\times10^{2}$& 6.5$\times10^{2}$& 0.72 & 8.5$\times10^{2}$& 1.2$\times10^{3}$& 0.68  \\
Q5a5 ($t=1.6\times 10^4$ yr) & 4.5$\times10^{2}$& 5.2$\times10^{2}$& 0.85 & 7.4$\times10^{2}$& 9.9$\times10^{2}$& 0.75  \\
Q5a8L ($t=2.3\times 10^4$ yr) & 3.2$\times10^{2}$& 5.2$\times10^{2}$& 0.62 & 6.9$\times10^{2}$& 9.9$\times10^{2}$& 0.69  \\
Q6a3 ($t=1.8\times 10^4$ yr)& 3.5$\times10^{2}$ & 4.3$\times10^{2}$& 0.79 & 7.3$\times10^{2}$& 8.2$\times10^{2}$& 0.88  \\
\enddata
\end{deluxetable*}

\subsection{Formation of Thin Dense Dust Rings}\label{subsec:dustyrings}
Figure \ref{fig:sigmaevolv_Q4a10ep01} shows time evolution of the dust and gas surface densities from Q4a10 run. We also show the surface density evolution from a run in which we switch off the self-gravity. We note that the gas disk is self-gravitationally stable and the dust GI is also stabilized because of the dust diffusion. We observe formation of multiple rings and gaps only when we switch on the self-gravity. Thus, the ring-gap formation results from the secular GI (see also Section \ref{subsec:linear}). The secular GI grows at wavelengths $\sim \cs/\Omega$, resulting in the ring-gap formation in the dust disk (see the dashed line in Figure \ref{fig:sigmaevolv_Q4a10ep01}). The resultant rings become much thinner, and the dust surface density increases. We do not observe significant substructures in the gas disk, which we discuss in more detail in Section \ref{sec:discussion}. 

On the left panel of Figure \ref{fig:t-r-sigma_Q4a10ep01}, we show the motion of dust cells that compose the resultant rings and gaps. The red dashed line shows the motions of the 525th cell as a reference cell composing one dust ring. We note that we reduced the number of cells to plot the left figure of Figure \ref{fig:t-r-sigma_Q4a10ep01}. Since our numerical scheme is based on the Lagrangian-cell method, the motion of cells represents actual motion of the dust, and a cell-concentrating region corresponds to a high density region. The dust initially moves inward with the steady drift velocity $v_{\mathrm{dri}}$ (Figure \ref{fig:vrQ3a5ep01}) 
\begin{equation}
v_{\mathrm{dri}}\equiv-\frac{2\tstop\Omega}{(1+\epsilon_0)^2+\left(\tstop\Omega\right)^2}\eta' r\Omega,\label{eq:2ddrift}
\end{equation}
where $\epsilon_0$ is the initial surface density ratio $\sigmad/\Sigma$ and $\eta'\equiv-\left(\cs^2/2r^2\Omega^2\right)\partial\ln\left(\cs^2\Sigma\right)/\partial\ln r$, and finally concentrates into the rings. One can see that the dust density perturbations also move inward with the drift velocity. The significant dust concentration ($t=1.3\times 10^5$ yr) results from the self-gravitational collapse of the dust rings that was also found in \citet[][]{Tominaga2018}. On the right panel of Figure \ref{fig:t-r-sigma_Q4a10ep01}, we plot the radial velocity at the dust cell boundaries ($r(t)=r_{\dst,i+1/2}$) that compose one dust ring ($i=520-530$). The position and the radial velocity of the 525th dust cell boundary is shown by the red solid line on the right panels. Those dust cells initially move inward with the drift velocity. As the secular GI grows ($r\gtrsim77$ au), the radial velocities deviate from the drift velocity and spread in the $v_r-r$ plane. When the cell width becomes smaller than the softening length as the dust surface density increases, the collapsing motion is suppressed (e.g., $r\simeq 77$au for $i=526-530$). Since the dust-to-gas surface density ratio gradually increases, the radial velocities of those cell boundaries decreases.

Figure \ref{fig:cumumass} shows the cumulative mass of the dust disks $M_{\dst}(<r)$. We can roughly evaluate the ring mass from Figure \ref{fig:cumumass} and find that the large amounts ($\sim$ tens of $\merth$) of dust grains reside in the individual ring. We define the radius of an individual dust ring as the position of the local maximum of the dust surface density, and estimate the mass of the $i$th ring $M_{\mathrm{ring},i}$ by summing up the mass of the dust cells between the adjacent local minima. We then compare the dust disk mass $M_{\dst}$ and the total ring mass $M_{\mathrm{ring,tot}}\equiv\sum_{i=1}^{N_{\mathrm{ring}}}M_{\mathrm{ring},i}$, where $N_{\mathrm{ring}}$ is the number of the dust rings whose amplitudes increase or are saturated at a certain time. In order to discuss the mass fraction of dust grains that are concentrated into rings, we set a certain radius denoted by $R_{\dst,\mathrm{out}}$ and use the dust cells whose initial radii are smaller than $R_{\dst,\mathrm{out}}$ to calculate $M_{\dst}$ and $M_{\mathrm{ring,tot}}$. The ring masses are calculated at the time when the dust-to-gas mass ratio in one of the resultant dust rings exceeds unity.

The results for $R_{\dst,\mathrm{out}}=120$ au and 200 au are summarized in Table \ref{tab:massfrac}. We find that the secular GI can convert tens percent of the dust mass into the ring mass. The ring mass $M_{\mathrm{ring,tot}}$ increases as increasing $R_{\dst,\mathrm{out}}$ since dust grains drifting from the outer disk are concentrated into rings at inner region (see Figure \ref{fig:t-r-sigma_Q4a10ep01}). 

The mass $M_{\dst}$ includes the mass of dust grains that are initially located at the secular-GI-stable region (see also Section \ref{subsec:linear}) and never concentrated into rings. Thus, the mass fraction $M_{\mathrm{ring,tot}}/M_{\dst}$ increases if one excludes those dust grains from the above estimation. Table \ref{tab:massfrac_ri60} summarizes $M_{\mathrm{ring,tot}},\;M_{\dst}$ and $M_{\mathrm{ring,tot}}/M_{\dst}$ estimated for dust grains initially located at $R_{\dst,\mathrm{in}}\leq r\leq R_{\dst,\mathrm{out}}$, where we set $R_{\dst,\mathrm{in}}=60\;\mathrm{au}$ since the secular GI can grow at $r>60\;\mathrm{au}$ in all runs. Over half of the dust masses are collected into the rings in most of the runs. Especially, 88 percent of the dust grains are saved in the dust rings in the case of Q6a3 run with $R_{\dst,\mathrm{in}}=60\;\mathrm{au}$ and $R_{\dst,\mathrm{out}}=200\;\mathrm{au}$.

The dust-to-gas mass ratio becomes comparable to or higher than unity as a consequence of the ring collapse, which makes the dust drift velocity smaller (see also the right panel of Figure \ref{fig:t-r-sigma_Q4a10ep01}). The increase in the dust density is saturated because of the balance between the diffusion and the self-gravitational collapse. Figure \ref{fig:vrdiffQ3a5ep01} compares two velocities around one dust rings whose radius is $r\simeq73.6$ au at $t=2.1\times10^4$ yr: (1) the dust velocity with respect to the ring velocity $v_{\mathrm{ring}}=-1.5\times10^{-3}$ au/yr that we measured from the data and (2) the velocity due to the dust diffusion $v_{\mathrm{diff}}\equiv-D\sigmad^{-1}\partial\sigmad/\partial r$. The red line shows the sum of them. We note that the red filled circles in the Figure \ref{fig:vrdiffQ3a5ep01} are the data points, which shows that the dusty ring is well resolved while the adjacent gaps are not. In the well-resolved ring, we find $|v_r-v_{\mathrm{ring}}|\simeq|v_{\mathrm{diff}}|>|v_r-v_{\mathrm{ring}}+v_{\mathrm{diff}}|$, which means that the dust diffusion quenches further self-gravitational collapse of the spiky dust rings. Especially, we can see the equilibrium at the inner half of the well-resolved ring. At the outer half of the ring, the red line shows increasing trend, which means that the radial converging speed is lower than the velocity due to the diffusion. This slow converging flow results from the deceleration due to the last term on the right hand side of Equation (\ref{eq:eomrdust}), i.e., $r^{-1}\partial F(r)/\partial r$ where $F(r)=rv_rD\partial\sigmad/\partial r$. Figure \ref{fig:forcecomp} compares four forces per mass acting on the dust: the self-gravity, the pressure gradient force ($\sigmad^{-1}\cd^2\partial\sigmad/\partial r$), the sum of the curvature term and the stellar gravity, and $\sigmad^{-1}r^{-1}\partial F(r)/\partial r$. The magnitude of the the force $\sigmad^{-1}r^{-1}\partial F(r)/\partial r$ is comparable to that of the self-gravity, meaning that the term decelerates the radial speed of the dust coming from the outer gap toward the ring.

Since we weaken the self-gravity with a constant softening length for the rings, the final dust surface density might be underestimated\footnote{The actual finite thickness of the disk slightly weakens the self-gravity for short-wavelength modes \citep[][]{Vandervoort1970,Shu1984}, and the softening term is expected to account for this effect to some extents. Thus, the following estimation of the dust surface density might be regarded as a reasonable upper limit.}. To check whether the dust density is underestimated or not, we evaluate a resultant dust surface density $\Sigma_{\dst,\mathrm{f}}$ which we would obtain if we neglect the softening. The diffusion timescale becomes comparable to the self-gravitational-collapse timescale after the nonlinear growth:
\begin{equation}
D^2k_{\mathrm{c}}^4\sim 2\pi G\Sigma_{\dst,\mathrm{f}} k_{\mathrm{c}},\label{eq:tdifftgi}
\end{equation}
where $k_{\mathrm{c}}^{-1}$ represents the length scale of the spiky ring. This gives $k_{\mathrm{c}}\sim\left(2\pi G\Sigma_{\dst,\mathrm{f}}/D^2\right)^{1/3}$. Assuming that the ring mass does not change throughout the linear and nonlinear growth, one has a relation between the surface density and the wavenumber: $\Sigma_{\dst,\mathrm{f}}/k_{\mathrm{c}}=\Sigma_{\dst,0}/k_0$, where $\Sigma_{\dst,0}$ and $k_0$ denote the unperturbed dust surface density and the wavenumber at which the secular GI grows in the linear phase, respectively. Using this relation and Equation (\ref{eq:tdifftgi}), one obtains
\begin{align}
\frac{\Sigma_{\dst,\mathrm{f}}}{\Sigma_{\dst,0}}&=\left(\frac{2\pi G\Sigma_{\dst,0}}{D^2}\right)^{\frac{1}{2}}k_{0}^{-\frac{3}{2}}\notag\\
&\simeq 9.3\left(\frac{\Sigma_{\dst,0}/\Sigma_{0}}{0.1}\right)^{\frac{1}{2}}\left(\frac{\alpha}{1\times10^{-3}}\right)^{-1}\left(\frac{Q}{4.5}\right)^{-\frac{1}{2}}\left(\frac{k_0H}{8}\right)^{-\frac{3}{2}},\label{eq:sigmadf}
\end{align}
where $\Sigma_0$ is the unperturbed gas surface density. This estimation is in good agreement with the resultant dust density of the ring at $r\simeq73.6$ au in Q4a10 run (see Figure \ref{fig:sigmaevolv_Q4a10ep01}). When the assumed $\alpha$ is smaller, numerical simulations with the constant softening length would underestimate $\Sigma_{\dst,\mathrm{f}}$.

\subsection{Formation of Transient Low-Contrast Dust Rings}\label{subsec:transientrings}
\begin{figure*}
	\begin{tabular}{c}
		\begin{minipage}{0.5\hsize}
			\begin{center}
				\includegraphics[width=1.0\columnwidth]{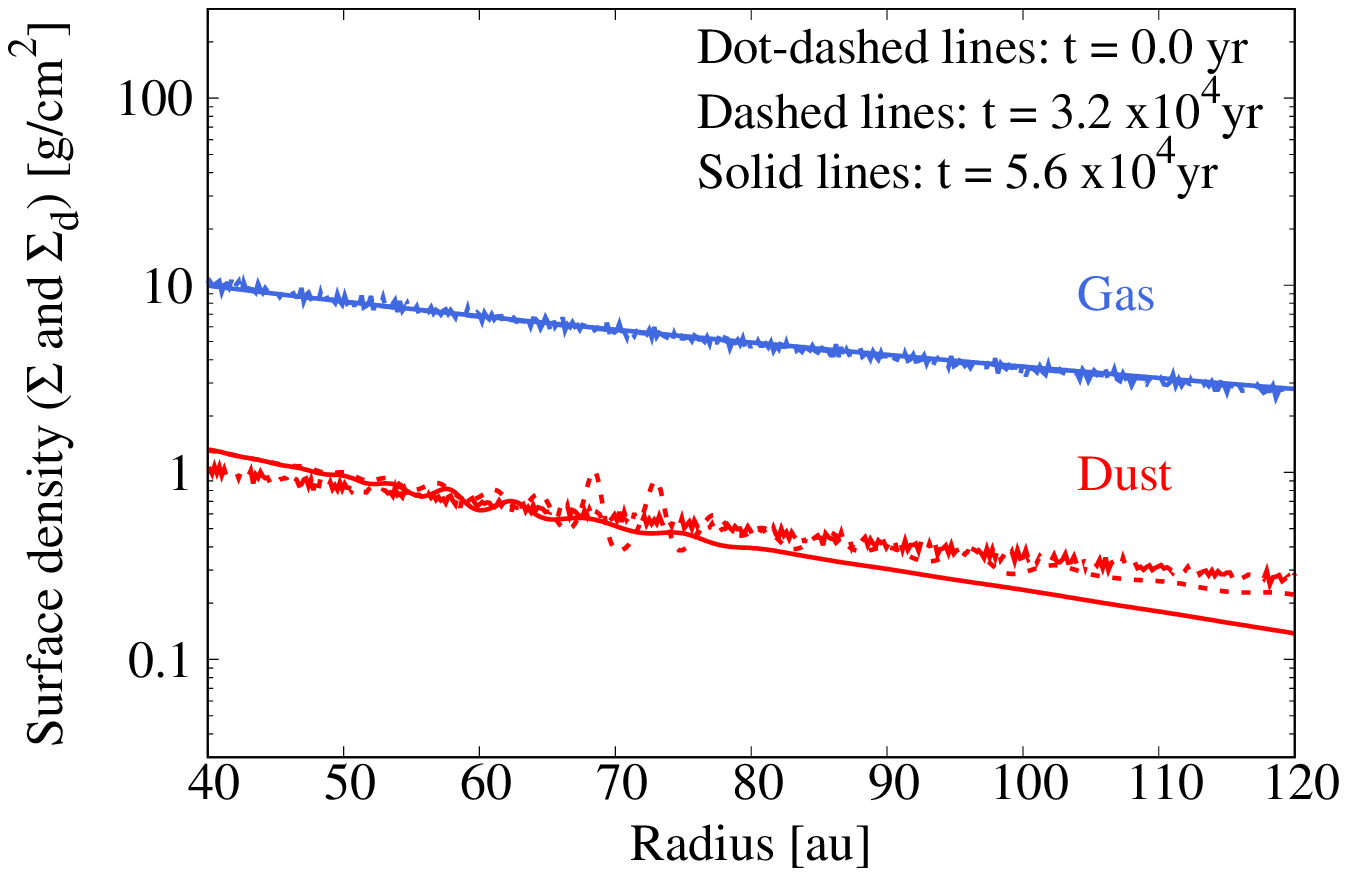}
			\end{center}
		\end{minipage}
		\begin{minipage}{0.5\hsize}
			\begin{center}
				\includegraphics[width=1\columnwidth]{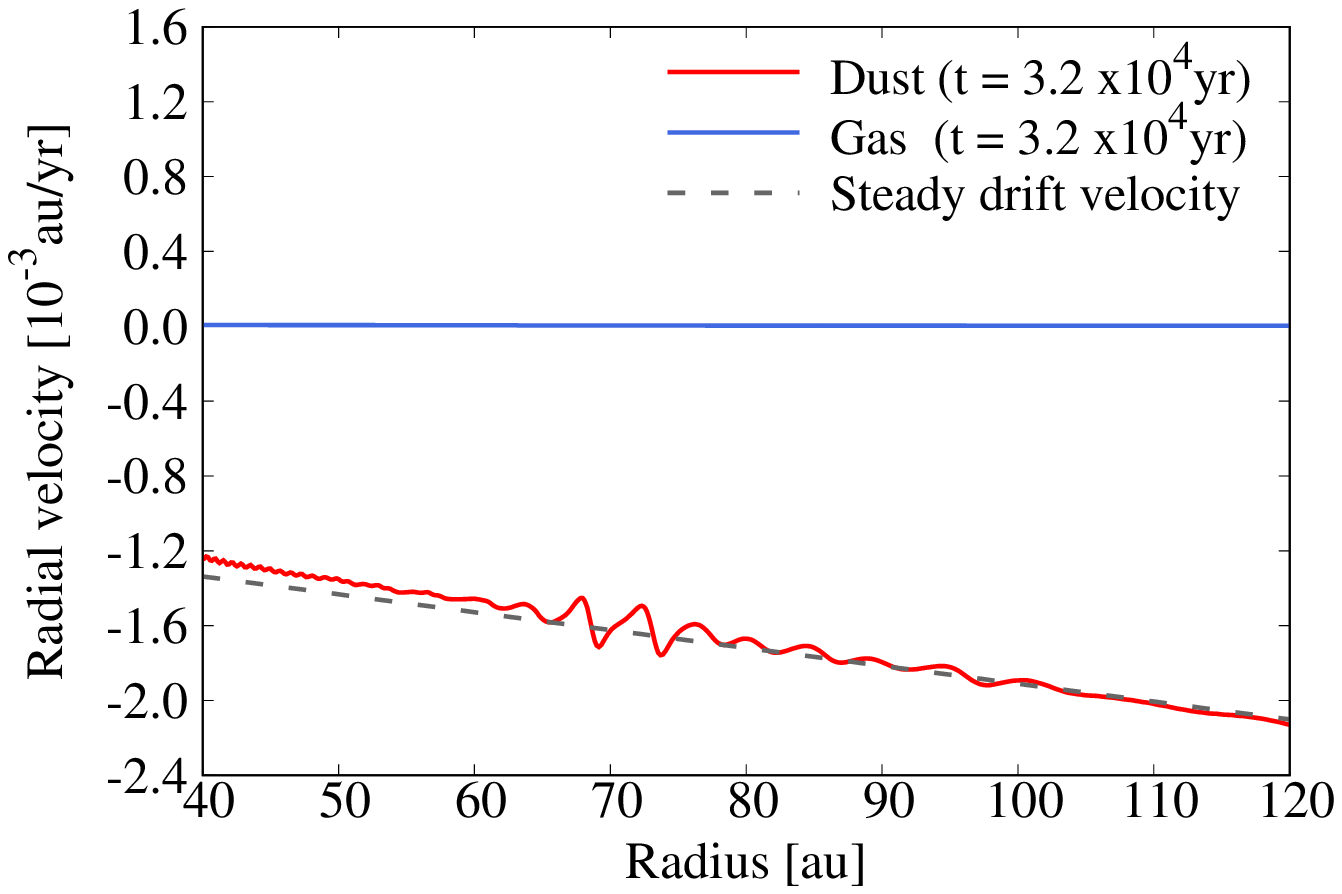}
			\end{center}
		\end{minipage}
	\end{tabular}
\caption{(Left panel) Surface density profile at $t=$0.0 yr (dot-dashed line), $3.2\times10^4$ yr (dashed line) and $5.6\times10^4$ yr (solid line) from Q5a8 run. The dust rings and gaps form with low-contrast. Those substructures decay with drifting inward. (Right panel) Radial velocity profile at $t=3.2\times10^4$ yr. The red and blue lines show the dust and gas radial velocities, respectively. The gray line is the steady drift velocity of dust \citep[][]{Nakagawa1986}. The mean radial velocity of the dust is in good agreement with the steady drift velocity as in Figure \ref{fig:vrQ3a5ep01} (Q4a10 run). In contrast, the relative velocity with respect to the drift motion is smaller. }
 \label{fig:densVrQ4a3ep01}
\end{figure*}
\begin{figure}[htp]
	\begin{center}
		\includegraphics[width=1.05\columnwidth]{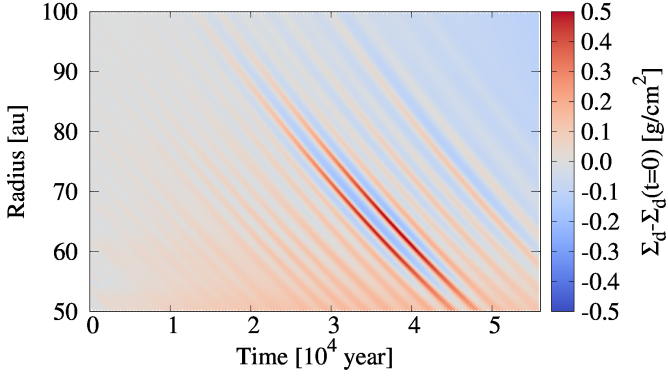}
	\end{center}
    		\caption{Deviation of dust surface density $\sigmad$ from the initial value $\sigmad(t=0)$as a function of radius and time in Q5a8 run. We clearly see the radial perturbations that move inward with time. Those become faint after they enter an inner region $r\lesssim 65$ au.} 
   		 \label{fig:dsigmamap_Q4a3ep01}
\end{figure}

Figure \ref{fig:densVrQ4a3ep01} shows the surface density and radial velocity profiles from Q5a8 run. We can see formation of dust rings and gaps at $t=3.2\times10^4$ yr, but these structures decay as they move inward, which is in contrast to Q4a10 run (Figure \ref{fig:sigmaevolv_Q4a10ep01}). The mean radial velocity of dust is in good agreement with the steady drift velocity as in the other runs. The relative velocity with respect to the drift motion is so low that significant dust concentration does not occur. As a result, the dust rings drift further without significant deceleration due to an increase of dust-to-gas mass ratio. 

The decay of the substructures in the dust surface density profile can be clearly seen in Figure \ref{fig:dsigmamap_Q4a3ep01} where we show deviation of $\sigmad$ from the initial profile $\sigmad(t=0)$ as a function of radius and time. Perturbations with a wavelength $\sim5$ au grow with moving inward from $r\gtrsim100$ au. The amplitudes decrease after they cross the radius $r\simeq60\mbox{--}70$ au. We also observe formation of those transient low-contrast rings in Q4a20 and Q6a5 runs. In these cases, the unstable region is so small that the dust moves across the region without significant increase in the surface density. We discuss this point in detail in the next section.

\section{Discussions}\label{sec:discussion}
We perform linear analyses of the secular GI including the drift motion of dust grains in order to obtain further understanding of the results of our simulations. In Sections \ref{subsec:linear} and \ref{subsec:condition}, we show the results of the linear analyses and discuss mode properties and a condition for significant dust concentration via the instability. In Section \ref{subsec:ringcoll}, we discuss subsequent ring evolution and implications on observed substructures. In Sections \ref{subsec:SI} and \ref{subsec:ddependD}, we mention two physical processes that are not included in the present simulations: streaming instability and density dependent diffusivity.

\subsection{Linear Stability Analyses and Mode Properties}\label{subsec:linear}
In this subsection, we derive the dispersion relation of the secular GI in the presence of the drift motion in order to understand the physics captured in the present simulations. We also show what kind of mode becomes the secular GI in the dust-drifting system. Although we consider both dust and gas in the present simulations, we start with linear analyses of one-fluid system to readily understand mode properties of the secular GI. This simplified analysis seems valid even for comparison with two-component simulations since the gas disk does not significantly evolve in our simulations\footnote{This does not mean that the back reaction on gas is unimportant. Indeed, the back reaction completely stabilizes the disk especially for long-wavelength modes as shown by \citet[][]{Takahashi2014}. In the case that the secular GI develops significantly, however, the contribution from the back reaction can be small, which enables our simplified analysis in this section.}. The two-fluid analysis is described later in this section.

We use the following equations that govern the dust motion in the local coordinates $(x,y)$ rotating around the central star with the angular velocity $\Omega_0=\Omega(r=R)$:
\begin{equation}
\frac{\partial\sigmad}{\partial t}+\frac{\partial}{\partial x}\left(\sigmad v_x\right)=D\frac{\partial ^2\sigmad}{\partial x^2},\label{eq:dusteoc_xy}
\end{equation}
\begin{align}
\frac{\partial v_x}{\partial t}+\left(v_x-\frac{D}{\sigmad}\frac{\partial\sigmad}{\partial x}\right)\frac{\partial v_x}{\partial x}&=3\Omega_0^2 x+2\Omega_0 v_y \notag\\
&-\frac{\cd^2}{\sigmad}\frac{\partial\sigmad}{\partial x}-\frac{\partial \Phi}{\partial x}-\frac{v_x}{\tstop}\notag\\
&+\frac{1}{\sigmad}\frac{\partial}{\partial x}\left(v_x D\frac{\partial\sigmad}{\partial x}\right),\label{eq:dusteomx_xy}
\end{align}
\begin{align}
\frac{\partial v_y}{\partial t}+\left(v_x-\frac{D}{\sigmad}\frac{\partial\sigmad}{\partial x}\right)\frac{\partial v_y}{\partial x}=&-2\Omega_0 \left(v_x-\frac{D}{\sigmad}\frac{\partial \sigmad}{\partial x}\right)\notag\\
&-\frac{v_y-U_{\gas,0}}{\tstop},\label{eq:dusteomy_xy}
\end{align}
where $U_{\gas,0}\equiv-3\Omega_0 x/2- \eta'_0 R\Omega_0$ is a steady azimuthal velocity of the background gas disk, and $\eta'_0\equiv\eta'(r=R)$. We include the dust diffusion so that the momentum is conserved in the absence of the drag force \citep[see subsection \ref{sec:basiceq} and][]{Tominaga2019}. An unperturbed state is the steady drift solution, $v_{x,0}, v_{y,0}$, with uniform surface density $\sigmad=\Sigma_{\dst,0}$:
\begin{equation}
v_{x,0}=-\frac{2\tstop\Omega_0}{1+\left(\tstop\Omega_0\right)^2}\eta'_0 R\Omega_0,
\end{equation}
\begin{equation}
v_{y,0}=U_{\gas,0}+\frac{\left(\tstop\Omega_0\right)^2}{1+\left(\tstop\Omega_0\right)^2}\eta'_0 R\Omega_0.
\end{equation}
Assuming that the perturbations $\delta\sigmad,\;\delta v_x,\;\delta v_y$ are proportional to $\exp[ikx+nt]$, one can derive the following linearized equations:
\begin{equation} 
(n+ikv_{x,0})\delta\sigmad+ik\Sigma_{\dst,0}\delta v_x=-Dk^2\delta\sigmad,\label{eq:lineocdust}
\end{equation}
\begin{align}
(n+ikv_{x,0})\delta v_x=&2\Omega_0\delta v_y-ik\cd^2\frac{\delta\sigmad}{\Sigma_{\dst,0}}-ik\delta\Phi\notag\\
&-\frac{\delta v_x}{\tstop}-v_{x,0}Dk^2\frac{\delta\sigmad}{\Sigma_{\dst,0}},
\end{align}
\begin{equation}
(n+ikv_{x,0})\delta v_y=-\frac{\Omega_0}{2}\left(\delta v_x-ikD\frac{\delta\sigmad}{\Sigma_{\dst,0}}\right)-\frac{\delta v_y}{\tstop}.
\end{equation}
We also assume that the background gas is not perturbed and $\delta\Phi=-2\pi G\delta\sigmad/k$. From the above linearized equations, we obtain the following dispersion relation for $\gamma\equiv n+ikv_{x,0}$:
\begin{align}
\left(\gamma+\frac{1}{\tstop}\right)&\biggl[F_{\mathrm{DW}}(\gamma,k)+\gamma\left(\frac{1}{\tstop}+Dk^2\right)\notag\\
&+Dk^2\left(-ikv_{x,0}+\frac{1}{\tstop}\right)\biggl]=\frac{\Omega_0^2}{\tstop}\label{eq:1fluiddisp},
\end{align}
\begin{equation}
F_{\mathrm{DW}}(\gamma,k)\equiv\gamma^2+\Omega_0^2+\cd^2k^2-2\pi G\Sigma_{\dst,0}k.
\end{equation}

In the absence of the dust diffusion ($D=0$), Equation (\ref{eq:1fluiddisp}) is equivalent to Equation (22) of \citet[][]{Youdin2011} except for the softening term of the self-gravity due to the finite disk thickness. The softening term does not essentially change the mode properties, and thus we do not include it here. One of the solutions of Equation (\ref{eq:1fluiddisp}) with $D=0$ corresponds to the one-component secular GI studied in \citet[][]{Youdin2005a,Youdin2011}, which is denoted by $\gamma_{\mathrm{SGI}}$ in the following, and we obtain $n=-ikv_{x,0}+\gamma_{\mathrm{SGI}}$. We note that Equation (\ref{eq:1fluiddisp}) with $D=0$ is a cubic equation of $\gamma$ with real coefficients, and $\mathrm{Im}[\gamma_{\mathrm{SGI}}]=0$\footnote{This is because the other roots correspond to the density waves and are complex conjugates for the self-gravitationally stable dust disk.}. This is mathematically and physically expected since we can remove the drift motion via the Galilean transformation \citep[see also][]{Youdin2005a}. The phase velocity $\mathrm{Re}[n/(-ik)]$ is the drift velocity $v_{x,0}$. This is consistent with our numerical simulations (see Figures \ref{fig:vrQ3a5ep01} and \ref{fig:densVrQ4a3ep01}). As mentioned in \citet[][]{Youdin2005a,Youdin2011}, the secular GI originates from a neutral mode, referred to as a ``static mode" in \citet[][]{Tominaga2019}, i.e. $\gamma_{\mathrm{SGI}}\to0$ for $\tstop\Omega\to\infty$. Our dispersion relation also shows $n\to 0$ for $\tstop\Omega\to\infty$ because $v_{x,0}$ also becomes zero. 

The above statement does not qualitatively change in the presence of the dust diffusion due to the weak gas turbulence. The dust diffusion just suppresses the growth of the secular GI especially at short wavelengths. Although strong dust diffusion results in non-zero $\mathrm{Im}[\gamma_{\mathrm{SGI}}]$ at short wavelengths, the secular GI does not grow ($\mathrm{Re}[\gamma_{\mathrm{SGI}}]<0$) with those wavelengths. Equation (\ref{eq:1fluiddisp}) is different from the dispersion relation derived by \citet[][]{Youdin2011} because the diffusion modeling is different. As mentioned in \citet[][]{Tominaga2019}, the diffusion modeling adopted in \citet[][]{Youdin2011} unphysically changes the dust angular momentum while ours does not. Hence, our dispersion relation (Equation (\ref{eq:1fluiddisp})) describes the mode properties more precisely. For example, \citet[][]{Youdin2011} found mode coupling between the neutral mode and the GI mode (unstable density waves) due to the diffusion while Equation (\ref{eq:1fluiddisp}) does not show such a mode coupling \citep[see also][]{Tominaga2019}.

We also perform two-fluid analyses with inclusion of the drift motion. In contrast to the above one-fluid analyses without the dust diffusion, we can not remove the drift motion since the dust and the gas have different drift speed. We use the following equations for the gas that are similar to those used to analyze the streaming instability \citep[e.g.,][]{Youdin2005,Youdin2007,Jacquet2011,Umurhan2020,Chen2020}:
\begin{equation}
\frac{\partial\Sigma}{\partial t}+\frac{\partial}{\partial x}\left(\Sigma u_x\right)=0,
\end{equation}
\begin{align}
\frac{\partial u_x}{\partial t}+u_x\frac{\partial u_x}{\partial x}=&3\Omega_0^2x+2\Omega_0u_y+2\eta'_0 R\Omega_0^2-\frac{\cs^2}{\Sigma}\frac{\partial\Sigma}{\partial x}\notag\\
&-\frac{\partial\Phi}{\partial x}+\frac{1}{\Sigma}\frac{\partial}{\partial x}\left(\frac{4}{3}\Sigma\nu\frac{\partial u_x}{\partial x}\right)+\frac{\sigmad}{\Sigma}\frac{v_x-u_x}{\tstop},
\end{align}
\begin{equation}
\frac{\partial u_y}{\partial t}+u_x\frac{\partial u_y}{\partial x}=-2\Omega_0u_x+\frac{1}{\Sigma}\frac{\partial}{\partial x}\left(\Sigma\nu\frac{\partial u_y}{\partial x}\right)+\frac{\sigmad}{\Sigma}\frac{v_y-u_y}{\tstop}.
\end{equation}
The equations for the dust are the almost same as Equations (\ref{eq:dusteoc_xy})-(\ref{eq:dusteomy_xy}) except for replacing $U_{\gas,0}$ by $u_y$ in Equation (\ref{eq:dusteomy_xy}). 

As in the previous one-fluid analyses, the unperturbed state is the steady drift solution where the unperturbed surface densities, $\Sigma_{\dst,0}$ and $\Sigma_0$, are spatially constant and the unperturbed velocities, $v_{x,0},\;v_{y,0},\;u_{x,0}$ and $u_{y,0}$, are the following \citep[][]{Nakagawa1986}:
\begin{equation}
v_{x,0}=-\frac{2\tstop\Omega_0}{(1+\epsilon)^2+\left(\tstop\Omega\right)^2}\eta'_0 R\Omega_0,
\end{equation}
\begin{equation}
v_{y,0}=-\frac{3}{2}\Omega_0x-\left[1-\frac{\left(\tstop\Omega_0\right)^2}{(1+\epsilon)^2+\left(\tstop\Omega\right)^2}\right]\frac{\eta'_0 R\Omega_0}{1+\epsilon},
\end{equation}
\begin{equation}
u_{x,0}=\frac{2\tstop\Omega_0\epsilon}{(1+\epsilon)^2+\left(\tstop\Omega\right)^2}\eta'_0 R\Omega_0,
\end{equation}
\begin{equation}
u_{y,0}=-\frac{3}{2}\Omega_0x-\left[1+\frac{\left(\tstop\Omega_0\right)^2\epsilon}{(1+\epsilon)^2+\left(\tstop\Omega\right)^2}\right]\frac{\eta'_0 R\Omega_0}{1+\epsilon},
\end{equation}
where $\epsilon\equiv\Sigma_{\dst,0}/\Sigma_0$. We fix $\tstop$ to comparison with the numerical simulations\footnote{Even if we consider the gas density dependence of $\tstop$ when linearizing the equation, the mode properties do not change much. }. The linearized continuity equation for the dust is the same as Equation (\ref{eq:lineocdust}). The other linearized equations are as follows:
\begin{equation}
(n+iku_{x,0})\delta\Sigma+ik\Sigma_0\delta u_x=0,
\end{equation}
\begin{align}
(n+iku_{x,0})\delta u_x=&2\Omega_0\delta u_y-ik\cs^2\frac{\delta\Sigma}{\Sigma_0}-ik\delta\Phi-\frac{4}{3}\nu k^2\delta u_x\notag\\
&+\frac{\delta\sigmad}{\Sigma_0}\frac{v_{x,0}-u_{x,0}}{\tstop}-\epsilon\frac{\delta\Sigma}{\Sigma_0}\frac{v_{x,0}-u_{x,0}}{\tstop}\notag\\
&+\epsilon\frac{\delta v_{x}-\delta u_{x}}{\tstop},
\end{align}
\begin{align}
(n+iku_{x,0})\delta u_y=&-\frac{\Omega_0}{2}\delta u_x-\frac{3}{2}\Omega ik\nu\frac{\delta\Sigma}{\Sigma_0}-\nu k^2\delta u_y\notag\\
&+\frac{\delta\sigmad}{\Sigma_0}\frac{v_{y,0}-u_{y,0}}{\tstop}-\epsilon\frac{\delta\Sigma}{\Sigma_0}\frac{v_{y,0}-u_{y,0}}{\tstop}\notag\\
&+\epsilon\frac{\delta v_{y}-\delta u_{y}}{\tstop},
\end{align}
\begin{align}
(n+ikv_{x,0})\delta v_x=&2\Omega_0\delta v_y-ik\cd^2\frac{\delta\sigmad}{\Sigma_{\dst,0}}-ik\delta\Phi\notag\\
&-\frac{\delta v_x-\delta u_x}{\tstop}-v_{x,0}Dk^2\frac{\delta\sigmad}{\Sigma_{\dst,0}},
\end{align}
\begin{equation}
(n+ikv_{x,0})\delta v_y=-\frac{\Omega_0}{2}\left(\delta v_x-ikD\frac{\delta\sigmad}{\Sigma_{\dst,0}}\right)-\frac{\delta v_y-\delta u_y}{\tstop},
\end{equation}
where $\delta\Phi=-2\pi G(\delta\Sigma+\delta\sigmad)/k$.
\begin{figure*}
	\begin{tabular}{c}
		\begin{minipage}{0.5\hsize}
			\begin{center}
				\includegraphics[width=0.9\columnwidth]{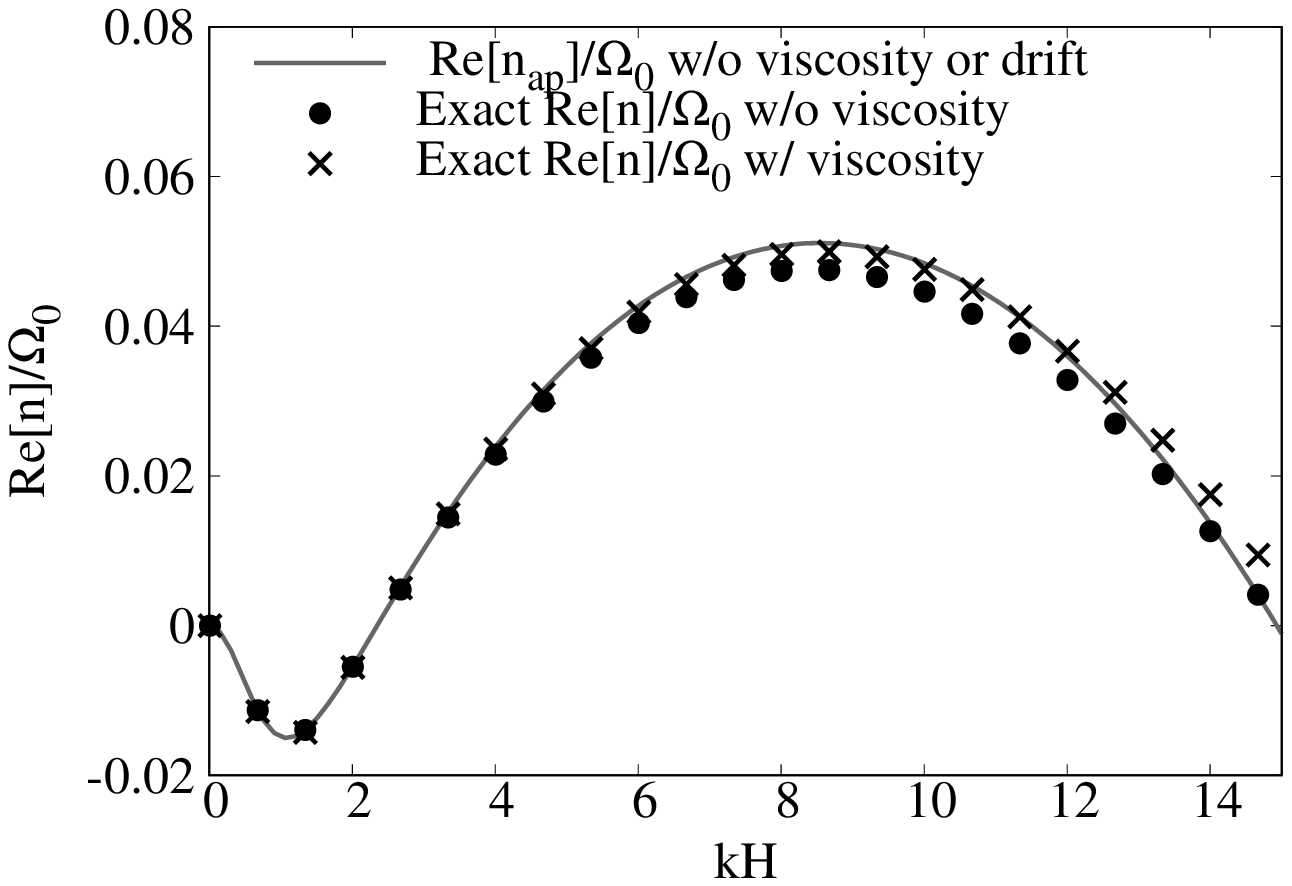}
			\end{center}
		\end{minipage}
		\begin{minipage}{0.5\hsize}
			\begin{center}
				\includegraphics[width=0.9\columnwidth]{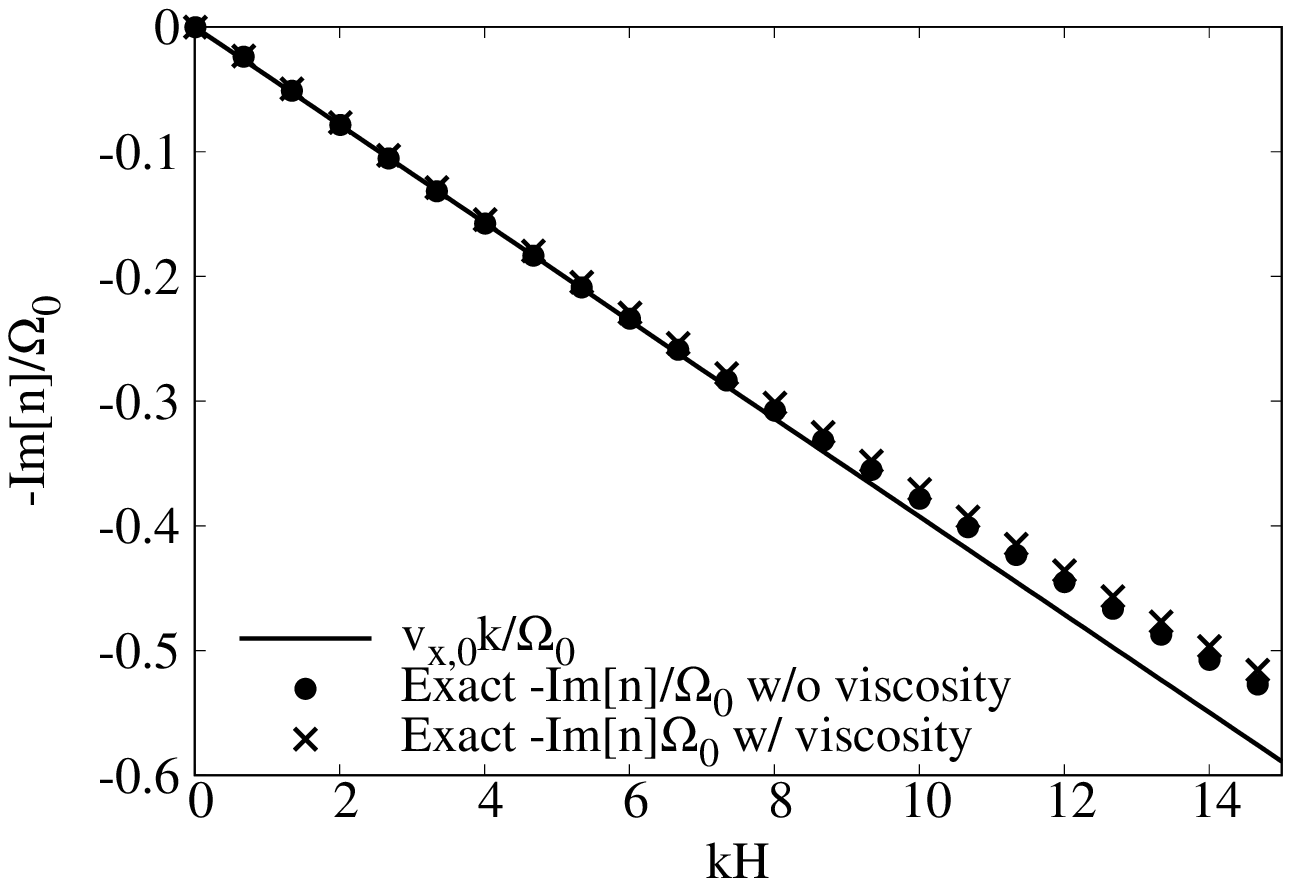}
			\end{center}
		\end{minipage}
	\end{tabular}
\caption{Dispersion relation of the secular GI at $r=75$ au for Q4a10 run. At the location of $r=75$ au, the values of physical parameters are $Q=4.463$ and $\eta'=0.003014$. The normalized stopping time and the dust-to-gas ratio are set to be 0.6 and 0.1, respectively. The normalized growth rate $\mathrm{Re}[n]/\Omega_0$ is shown on the left panel, and the normalized frequency $-\mathrm{Im}[n]/\Omega_0$ is on the right panel. The horizontal axis is the wavenumber normalized by the gas pressure scale height $H$. In both panels, the cross symbols and the filled circles show $n$ calculated with and without the turbulent viscosity, respectively. The gray line on the left panel shows the approximated dispersion relation in the case without the turbulent viscosity or the drift motion, which corresponds to Equation (26) of \citet[][]{Tominaga2019}. The black line on the right panel shows $v_{x,0}k/\Omega_0$ that reproduces well the oscillation frequency of the secular GI.}
 \label{fig:disp_Q3a5run}
\end{figure*}

In Figure \ref{fig:disp_Q3a5run}, we show the dispersion relation of the secular GI for Q4a10 run (cross marks). We set the unperturbed state using the physical values at $r=75$ au. The qualitative properties are the same as in the previous studies that neglected the drift motion. The long-wavelength perturbations are stabilized by the back reaction from dust to gas. The dust diffusion limits the shortest unstable wavelength. Thus, the secular GI is operational only at the intermediate wavelengths. 

The drift motion does not modify the growth rate of the secular GI significantly, which is similar to the above one-fluid analyses. To demonstrate this, we also show the growth rate without the turbulent viscosity (filled circles) with the approximated one $n_{\mathrm{ap}}$ that does not include the turbulent viscosity or the drift motion (gray line). The analytic expression of $n_{\mathrm{ap}}$ is given by \citet[][]{Tominaga2019} (see Equations (26)-(28) therein). The turbulent viscosity makes the growth rate larger (cross symbols) as also seen in the cases without the drift motion \citep[see,][]{Tominaga2019}. The frequency $-\mathrm{Im}[n]$ is shown on the right panel of Figure \ref{fig:disp_Q3a5run}. The frequency $v_{x,0}k$ (black line) well expresses the exact frequencies of the secular GI (filled circles and cross symbols) as in the one-fluid analyses (see also Appendix \ref{ap:qdepen}). At higher wavenumbers, the frequency deviates from $v_{x,0}k$ because of the dust diffusion term.

In the present analyses we do not find an unstable mode corresponding to TVGI discovered by \citet[][]{Tominaga2019}. In the previous analyses of \citet[][]{Tominaga2019}, TVGI originates from a static mode in which the dust velocity perturbations are in phase with the gas velocity perturbations. This situation is not realized if the drift velocity is significant, which seems to be the reason for the stabilization of TVGI. We address and analyze in more detail the effects of the drift motion on TVGI in our future work.

\begin{figure*}
	\begin{tabular}{c}
		\begin{minipage}{0.5\hsize}
			\begin{center}
				\includegraphics[width=0.9\columnwidth]{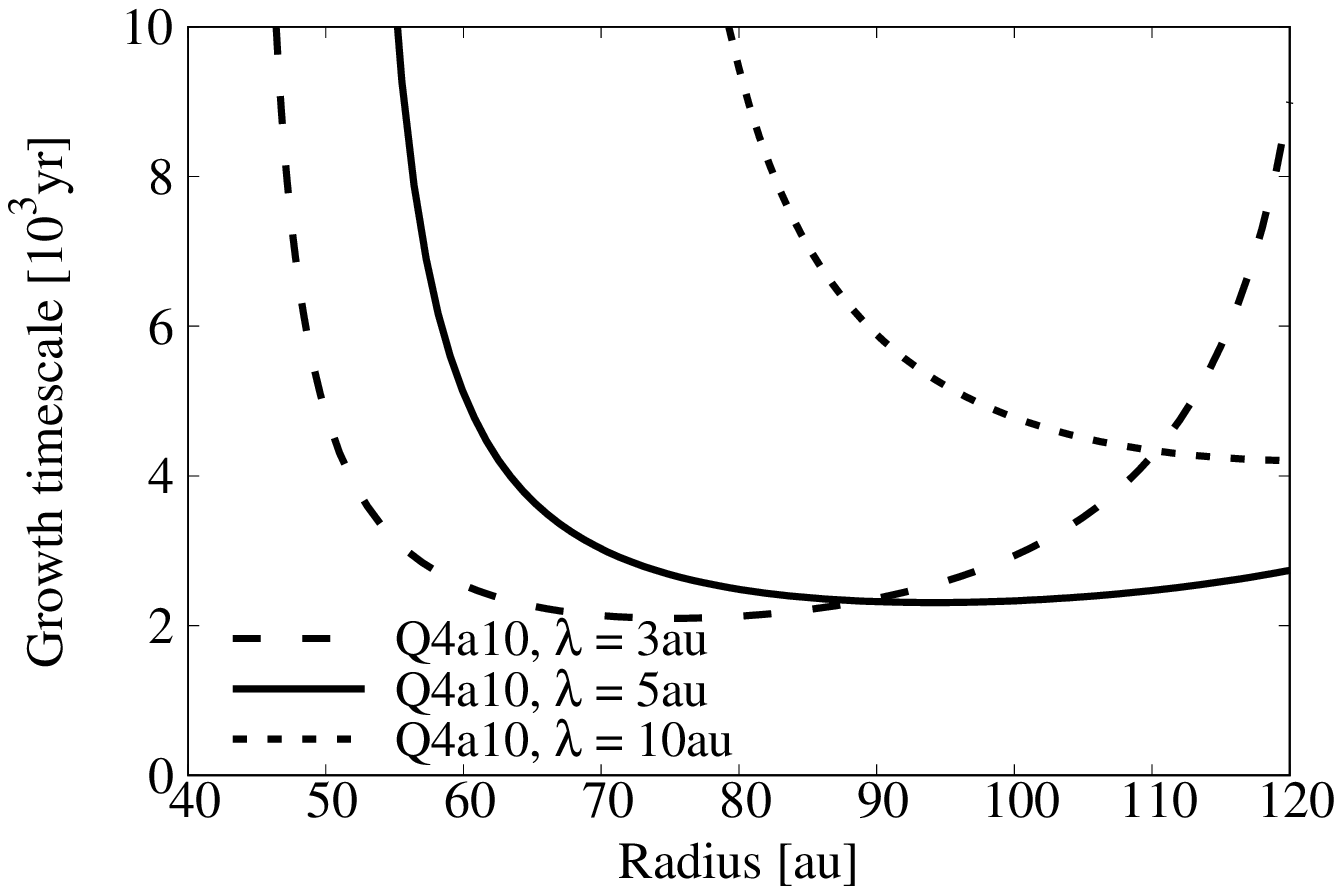}
			\end{center}
		\end{minipage}
		\begin{minipage}{0.5\hsize}
			\begin{center}
				\includegraphics[width=0.9\columnwidth]{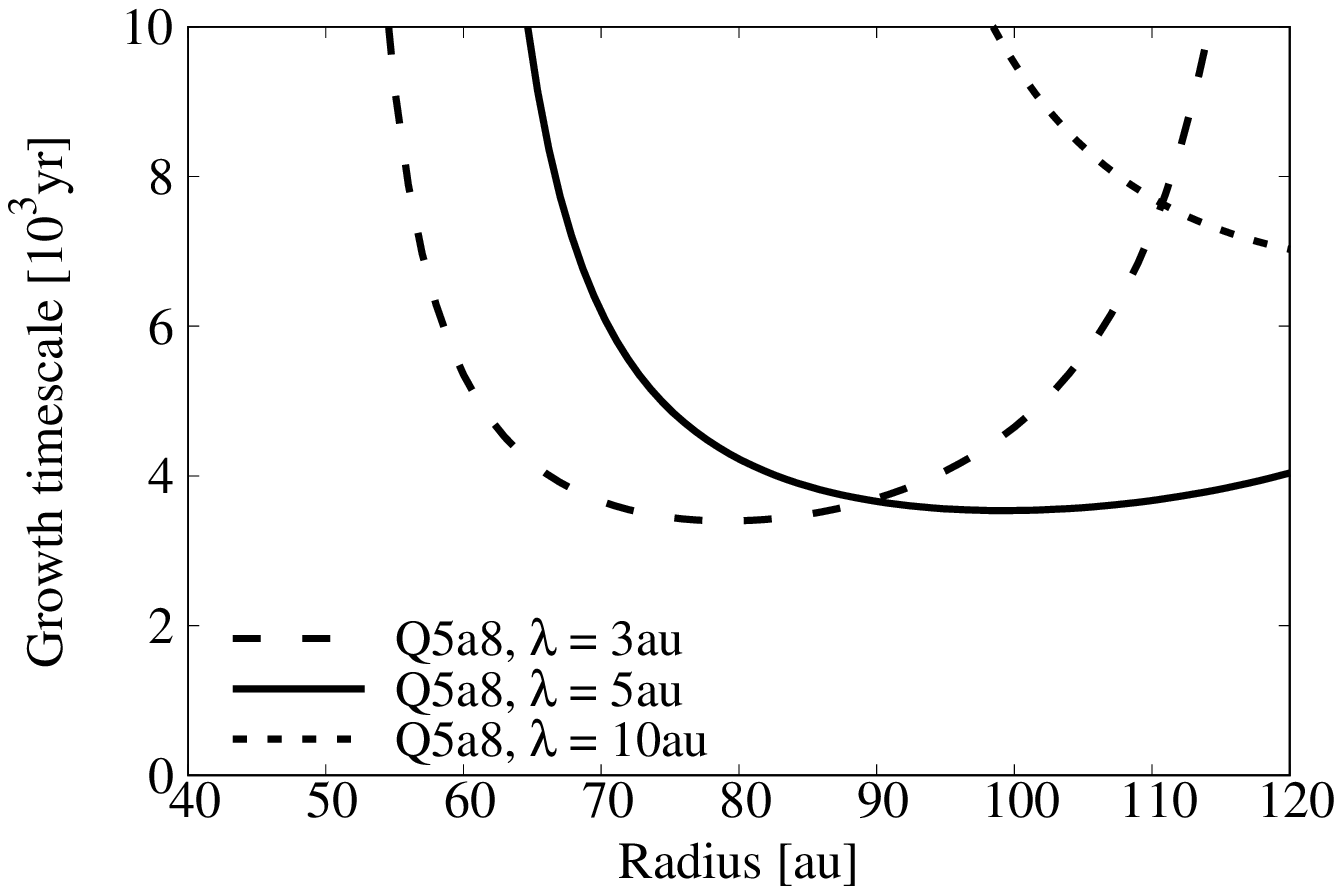}
			\end{center}
		\end{minipage}
	\end{tabular}
\caption{Growth timescale of the secular GI at wavelengths $\lambda=3\;\mathrm{au}, 5\;\mathrm{au}, 10\;\mathrm{au}$ in the cases of Q4a10 run (left panel) and Q5a8 run (right panel). The unstable region shifts outward in Q5a8 run with respect to that in Q4a10 run. Perturbations at long wavelength can grow at outer radii.}
 \label{fig:Q4a3tgrow}
\end{figure*}
\begin{figure}[htp]
	\begin{center}
		\includegraphics[width=1.0\columnwidth]{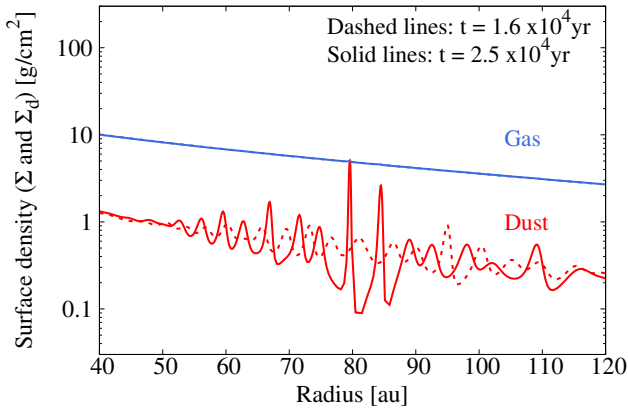}
	\end{center}
    		\caption{Surface density profile at $t=1.6\times10^4$ yr (dashed lines) and $2.5\times10^4$ yr (solid lines) from a run in which parameters are the same as Q5a8 run but the amplitude of the initial perturbations is six times larger. } 
   		 \label{fig:Q4a3-5largerpert}
\end{figure}
\begin{figure}[htp]
	\begin{center}
		\includegraphics[width=1.0\columnwidth]{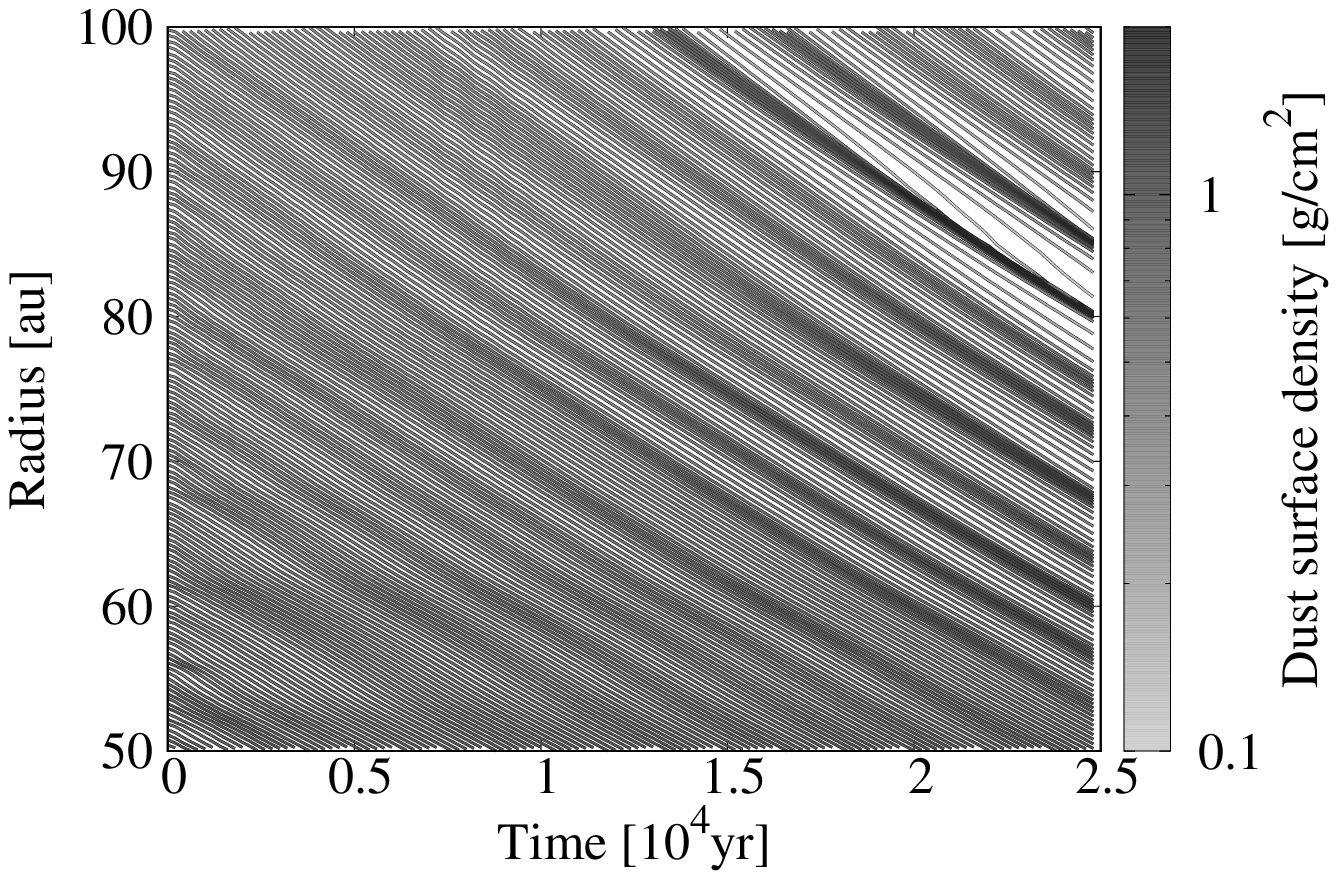}
	\end{center}
    		\caption{Time evolution of the dust-cell positions in Q5a8 run with six times larger initial perturbations. Color of the lines shows the dust surface density at each dust cell. We used a reduced number of cells in plotting this figure.} 
   		 \label{fig:t-r-sigma_Q4a3ep01-5largerpert}
\end{figure}

\subsection{Condition for the Thin Dense Ring Formation}\label{subsec:condition}
The condition for the formation of thin dense rings is that the radial drift motion is almost stopped because of the high dust-to-gas mass ratio in dust rings. Our numerical simulations indicate that the critical dust-to-gas mass ratio is about unity. In Q4a10 run, the maximum dust-to-gas mass ratio is about unity, and the thin dense rings form. On the other hand, the maximum dust-to-gas ratio is smaller than unity in Q5a8 run, and the transient low-contrast dust rings form (see Sections \ref{subsec:dustyrings} and \ref{subsec:transientrings}). We can understand the origin of these difference based on the linear stability analyses.
Figure \ref{fig:Q4a3tgrow} shows growth timescales of the secular GI in the cases of Q4a10 and Q5a8 runs. The wavelengths $\lambda$ are 3 au, 5 au and 10 au for each line. The minimum radii of the unstable regions for these wavelengths are larger for Q5a8 run than that for Q4a10 run. As shown in the previous section, the perturbations with $\lambda\simeq3-5$ au grow and form the substructures in both cases. In Q4a10 case, those perturbations can grow at $r\gtrsim50$ au, which is consistent with Figure \ref{fig:sigmaevolv_Q4a10ep01}. The drift velocity decelerates as the secular GI increases the dust-to-gas ratio, and the resultant rings tend to keep their radii once the dust-to-gas ratio approaches unity. As a result, the thin dense rings are not transient but sustain their dust-concentrating structure. On the other hand, in Q5a8 case the growth timescale at $\lambda=5$ au starts increasing significantly at $r\simeq60-70$ au. In other words, the growth of the perturbations decelerates, and the dust concentration gradually stops. Those perturbations keep drifting inward and eventually enter the stable region ($r\lesssim58$ au for $\lambda=5$ au). In the stable region, the perturbations decay and decrease their amplitudes as shown in Figure \ref{fig:dsigmamap_Q4a3ep01}.

The maximum dust-to-gas mass ratio in the rings does not only depend on the background disk parameter $Q$ and $\alpha$ but also depends on amplitudes of initial perturbations. We performed a simulation in which parameters are the same as Q5a8 run but the amplitude of the initial random perturbations is six times larger. We do not initially perturb a region of 10 au $<r<$ 20 au in order to avoid sudden dust concentrations due to the initial large perturbations near the inner boundary. Figure \ref{fig:Q4a3-5largerpert} shows the time evolution of the surface density. Rings with larger amplitudes form before they enter the stable region, resulting in thin dense rings with $\sigmad/\Sigma\simeq1$ (see also Figure \ref{fig:t-r-sigma_Q4a3ep01-5largerpert}). 
The amplitude of the initial perturbations in a real disk is unknown, and we do not discuss its origin in the present article.

In the present simulations, we fix the power-law index $q$ of the surface densities. As shown in the previous subsection and Appendix \ref{ap:qdepen}, $q$ insignificantly affect the growth timescale of the secular GI. In disks with larger $q$, the inner region is more massive than smaller-$q$ disks, and thus the unstable region shifts inward. If $\alpha,\;\tstop\Omega,\; Q$ and $\sigmad/\Sigma$ are the same in the shifted region, the growth timescale becomes shorter since $n/\Omega$ is constant for the same parameters and $n^{-1}\propto\Omega^{-1}\propto r^{3/2}$. The drift timescale also becomes shorter. The drift speed is faster for larger $q$. If one neglect the exponential cutoff term, $\eta' r\Omega$ is independent from the radial distance. Thus, the drift timescale is proportional to $r$ and becomes shorter in the shifted region. However, the $r$-dependence of the drift timescale is weaker than that of the growth timescale. Therefore, the secular GI can grow and more easily create thin dense rings in steeper disks. We note that the unstable wavelength becomes shorter since the gas scale height decreases inward. The widths of resultant rings are thus smaller than we present in Section \ref{sec:result}. 

\subsection{Subsequent Evolution and Observational Justification}\label{subsec:ringcoll}
The nonlinear growth of the secular GI results in dusty rings with $\sigmad\gtrsim\Sigma$. In such a massive dusty rings, 
the dust growth acceleration and planetesimal formation via the gravitational collapse are expected as described below.
Since the rings are massive with mass of $\gtrsim10\merth$ and self-gravitational, they will finally fragment in the azimuthal direction into solid bodies (see also Section \ref{subsec:ddependD}). The timescale of this ring fragmentation can be measured with the free-fall time $t_{\mathrm{ff}}$. The mean dust surface density in a dust ring $\tilde{\Sigma}_{\dst}$ is given by
\begin{equation}
\tilde{\Sigma}_{\dst}=\frac{M_{\ring}}{2\pi R_{\ring}w_{\ring}},\label{eq:sigmaestimate}
\end{equation}
where $M_{\ring},\; R_{\ring},\; w_{\ring}$ are mass, radius and width of a dust ring, respectively. We assume the mean mass density $\tilde{\rho}_{\dst}$ to be $\tilde{\Sigma}_{\dst}/\sqrt{2\pi}\hd$. The free-fall time is given by $t_{\mathrm{ff}}=\sqrt{3\pi/32G\tilde{\rho}_{\dst}}$. Taking the ring properties from Q4a10 run and assuming $M_{\ring}=49.7\merth$, $R_{\ring}=73.6$ au and $w_{\ring}=0.3$ au, we obtain $t_{\mathrm{ff}}\simeq55$ yr, which is much shorter than one Keplerian period at $r=73.6$ au ($\simeq631$ yr). 

Both the total dust-to-gas surface density ratio $\Sigma_{\dst,\mathrm{tot}}/\Sigma_{\mathrm{tot}}$ and the dust-to-gas ratio $\sigmad/\Sigma$ in the lower disk increases in the growth of the dusty rings. Since the collisional growth timescale of dust grains $t_{\mathrm{grow}}$ is inversely proportional to the total dust-to-gas surface density ratio $\Sigma_{\dst,\mathrm{tot}}/\Sigma_{\mathrm{tot}}$ in the Epstein regime (see Equation (\ref{eq:tgrowEp})), the dust growth should proceed faster in the dusty rings.
When the height of the lower layer is much larger than the dust scale height, one has $\Sigma_{\dst}\simeq\Sigma_{\dst,\mathrm{tot}}$. Since $\Sigma$ insignificantly changes during the growth of the secular GI, $\Sigma_{\mathrm{tot}}$ will be almost constant in time. Thus, the enhancement factor of $\Sigma_{\dst,\mathrm{tot}}/\Sigma_{\mathrm{tot}}$ will be similar to that of $\Sigma_{\dst}/\Sigma$.
In the case of Q4a10 run, if the dust-to-gas ratio $\Sigma_{\dst,\mathrm{tot}}/\Sigma_{\gas,\mathrm{tot}}$ is initially 0.05 and increases as well as $\sigmad/\Sigma$ by a factor of 9.3 as expected from the estimation of $\Sigma_{\dst,\mathrm{f}}/\Sigma_{\dst,0}$ (see Equation (\ref{eq:sigmadf})), Equation (\ref{eq:tgrowEp}) gives $t_{\mathrm{grow}}\simeq199$ yr at $r=73.6$ au, which is about 3.1 times smaller than one orbital period although it is still larger than $t_{\mathrm{ff}}$.

In the case of Q5a8 run with six times larger initial perturbations, $\Sigma_{\dst}/\Sigma$ becomes an order of magnitude larger than the initial value at $r\simeq80$ au (Figure \ref{fig:Q4a3-5largerpert}). The expected dust growth timescale in the dust ring is about 209 yr, which is much shorter than the radial drift timescale. Thus, the dust grains in the ring will grow and decouple from the gas before the dust ring drifts into the stable region, which will also prevent the ring from being transient.
Because we only include the turbulence-driven relative velocity in Equation (\ref{eq:tgrowEp}), the dust growth timescale is overestimated. Thus, to consider both ring fragmentation and dust growth is important, and in both cases, we can expect the subsequent planetesimal formation if the significant dust concentration occurs via the secular GI. More quantitative analyses require multidimensional simulations with the dust growth, which is our future work.

As mentioned in Section \ref{sec:result}, the secular GI results in significant substructures only in a dust disk. This property is in contrast to the planet-based scenario in which unseen Jupiter-mass planets carve gaps both dust and gas disks \citep[e.g.,][]{Gonzalez2015,Kanagawa2015,Zhang2018}. Therefore, observational determination of a midplane gas density profile would enable us 
to discriminate the ring formation mechanisms that require gap-like profile in the gaseous component (i.e., hidden high-mass planet hypothesis) and that shows relatively smooth profile in gas (i.e., the present mechanism).

We should note that if the nonlinear growth of the secular GI results in massive rocky objects, those objects would create some substructures in the gas disk. 
If the gas gap is as large as the dust gap, it might be difficult to clearly distinguish the secular-GI-based mechanism and the planet-based mechanism. Even in this case, however, the secular GI might explain the origin of the hypothetical planets in the gaps. 

Even if a gas gap is not observed around a dust gap, it is possible that a low-mass planet carves the dust gap. A low-mass planet first creates prominent structures in a dust disk and takes a long time to carve a gas gap since the dust scale height is smaller than the gas scale height \citep[e.g.,][]{Yang2020}. Thus, there is a possibility that observations only identify the resolvable dust gaps but cannot resolve low-contrast gas gaps. Thus, the low-mass-planet-based scenario and the secular-GI-based scenario can be degenerate. Nevertheless, those scenarios depend on the background density and pressure profiles, the dust drift and sizes in different ways. Therefore, it is important to accurately measure the gas density profile near the midplane as well as the dust surface density.

Separations and widths of rings/gaps depend on at which wavelengths perturbations are large and grow fast via the secular GI. In our simulations with the initial random perturbations, the wavelength is comparable to the gas scale height at each radius, which is about the most unstable wavelength. The rings migrate inward while keeping separations of the adjacent rings almost constant (Figure \ref{fig:dsigmamap_Q4a3ep01}) because the radial variation of the drift velocity is small within a spatial scale of the unstable wavelength. This is another feature of the ring-forming process via the secular GI. 

Thin dense rings forming via the secular GI may be observed as marginally optically thin rings although the solid surface density in those rings can be much high. \citet{Stammler2019} claims that planetesimal formation via the streaming instability in a pressure bump can explain the marginally optically thick rings \citep[see also,][]{Huang2018,Dullemond2018}. The streaming instability in a bump converts dust grains into planetesimals and stalls itself as a result of the dust depletion. Since the resultant planetesimals are not observed at sub-mm emission, this self-regulating process can limit the optical depth in the dust ring. Similar effects is expected in the ring formation via the secular GI.

As discussed above, dust growth to larger solid bodies can occur in the resultant dusty rings. If those solid bodies are large enough, those rings would be dark at millimeter wavelengths, and the multiple spiky rings and adjacent gaps would be observed as a wide gap structure. Collisional fragmentation of the resultant larger bodies would supply small grains. The remaining and resupplied small grains takes part in the growth of the secular GI and contribute to sub-mm emission observed by ALMA. To examine what kind of substructures are expected in the intensity profile, we have to explicitly include the dust growth and fragmentation in our simulations. Since self-gravitational fragmentation of the rings could occur along with the dust growth at comparable timescale, non-axisymmetic analyses and simulations are important to explore observational signatures. Those will be the scope of our future studies.

\subsection{Relation between the secular GI and the streaming instability}\label{subsec:SI}
In the presence of the dust drift motion, streaming instability can operate in a disk \citep[][]{Youdin2005,Youdin2007,Jacquet2011}. Nonlinear properties of the streaming instability have been extensively studied based on local-shearing-box simulations \citep[e.g.,][]{Johansen2007,Johansen2007nature,Johansen2009,Bai2010a,Bai2010b,Bai2010c,Yang2014,Simon2016,Yang2017,Schreiber2018}. These numerical studies showed that the streaming instability creates dust clumps and triggers GI if those clumps are enough massive. This self-gravitational clump collapse results in planetesimals. Thus, the self-gravity and GI are required to trigger planetesimal formation even in the scenario based on the streaming instability. 

In the presence of the self-gravity the secular GI can also operate as we show in the Section \ref{subsec:linear} and thus would be important to form dust clumps and planetesimals. \citet[][]{Abod2019} studied the streaming instability based on the local simulations with the self-gravity and showed dust clumping even in the absence of the global pressure gradient. The clumping they found is caused by the classical GI mediated by the dust-gas friction although they referred to the process as the secular GI. The secular GI is essentially different from the friction-mediated GI \citep[][]{Youdin2005,Tominaga2019}, and thus the relation between the secular GI and the streaming instability is still unclear.

Our linear analyses and nonlinear simulations show that the secular GI grows at wavelengths $\sim H$ for $\alpha\sim10^{-3}$ and $\tstop\Omega=0.6$. The streaming instability grows at shorter wavelengths $\ll H$ \citep[e.g., see][]{Youdin2005} and creates smaller structures. Therefore, the secular GI and the streaming instability would grow independently at each scale at least in the linear growth phase.

In the presence of the dust diffusion as in our study, the growth of the streaming instability would be suppressed \citep[][]{Umurhan2020,Chen2020}, and the dust diffusion limits spatial scales at which subsequent gravitational collapse can proceed \citep[][]{Gerbig2020,Gole2020,Klahr2020}. The linear or nonlinear growth of the secular GI will potentially provide a dust-rich environment that is necessary for the streaming instability to grow against the diffusion. Therefore, the combination of dust concentration at large scale ($\sim H$) due to the secular GI and further concentration at small scale ($\ll H$) due to the streaming instability is one possible path to form planetesimals.

Depending on the driving mechanism of gas turbulence, the radial diffusion is inefficient in contrast to the vertical diffusion \citep[][]{Yang2018,Schafer2020}. In this case, both secular GI and streaming instability can relatively easily develop against the diffusion.
To explore more detailed mode properties, nonlinear coupling, and evolutionary paths toward planetesimal formation, multidimensional linear analyses and simulations with the radial extent of $\sim H$ are necessary. These will also be the scope of our future work.

\subsection{Effects of the dust-to-gas mass ratio dependence of the dust diffusivity}\label{subsec:ddependD}
In our simulations, the dust diffusivity $D$ is given by a constant parameter $\alpha$. However, it will decrease when the large dust-to-gas mass ratio is realized in the rings formed by the secular GI (see Figures \ref{fig:sigmaevolv_Q4a10ep01}). For example, \cite{Schreiber2018} has reported that the turbulent diffusion caused by the streaming instability weakens when the dust-to-gas mass ratio is larger than unity. If the turbulent diffusion in dust rings is driven by the streaming instability, the saturated surface density of the thin dense rings will be larger than that obtained from our simulations or estimated in Equation (\ref{eq:sigmadf}). It is also possible that the inefficient diffusion cannot support the thin dense ring and the ring just radially collapses. In either case, the planetesimal formation timescale via the gravitational collapse or the dust growth will be shorter than that estimated in the previous section. 

We can estimate the saturated surface density at dense thin rings as follows. We assume that the diffusion coefficient of dust is given by
\begin{equation}
D=\alpha_1\frac{\cs^2}{\Omega}\left(\frac{\Sigma_{\dst,\mathrm{f}}}{\Sigma_{0}}\right)^{-\beta},
\end{equation}
where $\alpha_1$ is dimensionless diffusion parameter when $\Sigma_{\dst,\mathrm{f}}/\Sigma_0=1$. In the same way to derive Equation (\ref{eq:sigmadf}), we can evaluate the dust surface density when it achieves saturation as follows:
\begin{align}
\frac{\Sigma_{\dst,\mathrm{f}}}{\Sigma_{\dst,0}}=&f(\beta)\left(\frac{\Sigma_{\dst,0}/\Sigma_{0}}{0.1}\right)^{\frac{1+2\beta}{2-2\beta}}\left(\frac{\alpha_1}{1\times10^{-3}}\right)^{\frac{1}{\beta-1}}\notag\\
&\times\left(\frac{Q}{4.5}\right)^{\frac{1}{2\beta-2}}\left(\frac{k_0H}{8}\right)^{\frac{3}{2\beta-2}},\label{eq:sigmadf_modi}
\end{align}
\begin{equation}
f(\beta)\simeq(8.7\times 10^{1-2\beta})^{\frac{1}{2-2\beta}}.
\end{equation}
The equation has no solution and $f(\beta)$ diverges if $\beta=1$. This indicates that the dust diffusion can not sustain the gravitational collapse of the dust ring if the dust-to-gas mass ratio dependence of $D$ is steeper than the case of $\beta=1$. Although the surface density has a solution for $\beta>1$, it is not achieved during the growth of the perturbation. In the case of $\beta>1$, the diffusion timescale is smaller than the collapse timescale when the dust ring surface density is smaller than $\Sigma_{\dst,\mathrm{f}}$. 

The velocity dispersion of dust grains $\cd$ is expected to decrease as the dust-to-gas ratio increases beyond unity. If the decrease of $\cd$ is realized, the Coriolis force will be a unique repulsing process against the self-gravitational collapse. The characteristic lengthscale $\lambda_{\mathrm{crit}}$ of such a system is
\begin{equation}
\lambda_{\mathrm{crit}}\equiv\frac{4\pi^2 G\sigmad}{\Omega^2}.
\end{equation}
The dust clumps smaller than $\lambda_{\mathrm{crit}}$ will self-gravitationally collapse. Taking the ring properties from Q4a10 run and adopting $\sigmad=9.5$ $\mathrm{g/cm}^2$ at $r=73.6$ au, one obtains $\lambda_{\mathrm{crit}}\simeq17.7$ au, which is much larger than the resultant ring width (see Figure \ref{fig:sigmaevolv_Q4a10ep01}). The circumference of a ring with the radius of 73.6 au is about 462 au, which is 6 times longer than $\lambda_{\mathrm{crit}}$. Thus, non-axisymmetric modes with an azimuthal wavenumber $m$ larger than 6 might grow and cause azimuthal fragmentation. Since higher $m$ modes have larger growth rate, sizes of resultant dust clumps can be much smaller. If the dust ring in Q4a10 run whose mass is $M_{\mathrm{ring}}=49.7\merth$ and radius is $r=\simeq73.6$ au fragments, the mass of the dust clumps $M_{\mathrm{clump}}$ resulting from the azimuthal fragmentation is 
\begin{equation}
M_{\mathrm{clump}}=\frac{M_{\mathrm{ring}}}{m}\simeq8.2\merth\left(\frac{m}{6}\right)^{-1}.
\end{equation}
Since the non-axisymmetric modes with $m<6$ are expected to be stable, the above clump mass for $m=6$ gives upper limit. For further studies, non-axisymmetric simulations are necessary.

\section{Conclusions}\label{sec:conclusion}
In this work, we investigate the linear/nonlinear growth of the secular GI with explicitly including the dust drift for the first time. Our numerical simulations show that the secular GI can operate even when dust grains drift inward and create dusty rings, which we can understand the linear analyses including the drift motion. 

The nonlinear growth can increase the dust surface density by an order of magnitude. The drift motion is suppressed in the resultant rings once the dust-to-gas ratio exceeds unity, and thin dense rings form. If the growth timescale of the secular GI is too long and the dust-to-gas ratio remains smaller than unity, the resultant substructures become transient and start decaying once they enter the inner region stable to the instability. 

The simple estimates of the dust-growth timescale and the free-fall time show the possibility of the subsequent planetesimal formation through the fast grain growth or the self-gravitational ring fragmentation. When the ring structures formed by the secular GI achieve the nonlinear growth, observable high contrast dusty ring structures and planetesimals form. Although the dusty rings decay in the inner stable region, the ring-gap contrast smoothly decreases as the rings move inward. Thus, it seems possible that rings and gaps are observed even in the inner stable region. 

For further investigation of the planetesimal formation and observational predictions of the secular GI, we have to consider the dust grain growth and the non-axisymmetric ring fragmentation. In contrast to the planet-disk interaction scenario, the secular GI does not create significant gas substructures. Therefore, gas observations with optically thin lines that can trace the midplane gas density profile will provide important insight to understand what kind of mechanisms actually operate. 

\acknowledgments
We thank the anonymous referee for detailed review and insightful comments that helped to improve the manuscript. We also thank H. Kobayashi and H. Tanaka for fruitful comments and discussions. This work is supported by JSPS KAKENHI Grant Number JP18J20360, 18H05438, and 19K14764.

%




\appendix

\section{Drift-Limited Stopping Time}\label{ap:tstop}

In this appendix, we derive the drift-limited stopping time using our disk model. The drift-limited stopping time in different disk models was derived in the previous work \citep[e.g.,][]{Birnstiel2012,Okuzumi2012}. Following the previous work, we compare the two timescales: dust growth timescale $t_{\mathrm{grow}}$ within which the dust size becomes twice larger, and drift timescale $t_{\drift}$ within which the dust drifts and falls onto the central star.

The dust growth timescale for spherical grains is given by
\begin{equation}
t_{\mathrm{grow}}=3m\left(\frac{dm}{dt}\right)^{-1}=\frac{3m}{\rho_{\dst}\sigma\Delta v},\label{eq:tgrowpre}
\end{equation}
where $m$ and $\rho_{\dst}$ are mass of a single dust grain and the density of the dust disk. The cross section and the relative velocity of the dust grains are denoted by $\sigma$ and $\Delta v$, respectively. For compact spherical dust grains with the radius of $a$ and the internal density of $\rho_{\mathrm{int}}$, Equation (\ref{eq:tgrowpre}) is reduced to
\begin{equation}
t_{\mathrm{grow}} = \frac{\rho_{\mathrm{int}}a}{\rho_{\dst}\Delta v}. \label{eq:tgrow}
\end{equation}
We consider turbulence-driven relative velocity $\Delta v=\sqrt{3\tstop\Omega\alpha}\cs$ \citep[][]{Ormel2007} and the Epstein drag regime
\begin{equation}
\tstop=\sqrt{\frac{\pi}{8}}\frac{\rho_{\mathrm{int}}a}{\rho_{\gas}\cs},\label{eq:epstein}
\end{equation}
where $\rho_{\gas}$ is the gas density. The vertical dust and gas density profiles are assumed to follow the Gaussian function:
\begin{equation}
\rho_{\gas}=\frac{\Sigma_{\mathrm{tot}}}{\sqrt{2\pi}H}\exp\left(-\frac{z^2}{2H^2}\right),\label{eq:rhogas}
\end{equation}
\begin{equation}
\rho_{\dst}=\frac{\Sigma_{\dst,\mathrm{tot}}}{\sqrt{2\pi}H_{\dst}}\exp\left(-\frac{z^2}{2H_{\dst}^2}\right),\label{eq:rhodst}
\end{equation}
where $H_{\dst}\simeq H\sqrt{\alpha/\tstop\Omega}$ is the dust scale height \citep[e.g.,][]{Dubrulle1995,Carballido2006,YL2007}.  Equations (\ref{eq:tgrow})-(\ref{eq:rhodst}) give the dust growth timescale at the midplane ($z=0$):
\begin{equation}
t_{\mathrm{grow}}\Omega\simeq2\sqrt{\frac{2}{3\pi}}\frac{\Sigma_{\mathrm{tot}}}{\Sigma_{\dst,\mathrm{tot}}}.\label{eq:tgrowEp}
\end{equation}
Note that the growth timescale depends not on $\sigmad/\Sigma$ but on $\Sigma_{\dst,\mathrm{tot}}/\Sigma_{\mathrm{tot}}$.

The drift timescale $t_{\drift}\Omega$ is given by
\begin{equation}
t_{\drift}\Omega=\frac{r\Omega}{|v_{r,\drift}|}.\label{eq:tdrift}
\end{equation}
The denominator $v_{r,\drift}$ is the steady drift velocity \citep[][]{Nakagawa1986}
\begin{equation}
v_{r,\drift}=-\frac{2\tstop\Omega}{\left(1+\varepsilon\right)^2+\left(\tstop\Omega\right)^2}\eta r\Omega,\label{eq:vdrift}
\end{equation}
where $\varepsilon$ is the local dust-to-gas mass ratio $\rho_{\dst}/\rho_{\gas}$ and 
\begin{equation}
\eta=-\frac{1}{2\rho_{\gas}r\Omega^2}\frac{\partial \left(\cs^2\rho_{\gas}\right)}{\partial r}=-\frac{H^2}{2r^2}\frac{\partial\ln\left(\cs^2\rho_{\gas}\right)}{\partial\ln r}.\label{eq:eta}
\end{equation}
Assuming that the radial profile of $\Sigma_{\mathrm{tot}}$ has the same power law index $q$ as $\Sigma$ (see Equation (\ref{eq:gasdiskmodel})), Equation (\ref{eq:eta}) at the midplane ($\rho_{\gas}(r,z=0)=\Sigma_{\mathrm{tot}}(r)/\sqrt{2\pi}H(r)$) becomes\footnote{We should note that $\eta$ given by Equation (\ref{eq:eta}) is generally different from $\eta'\equiv-\left(\cs^2/2r^2\Omega^2\right)\partial\ln\left(\cs^2\Sigma\right)/\partial\ln r$, which determines the dust drift velocity in the vertically integrated system (see also Equation (\ref{eq:2ddrift})). The difference comes from (1) the radial variation of $H(r)$ and (2) the fact that $\Sigma$ is different from $\Sigma_{\mathrm{tot}}$ (see Section \ref{sec:basiceq}). For example, the former gives $\eta=\eta'(7/4+q)/(1/2+q)$ if we neglect the exponential cutoff of the gas surface density. The difference from the latter depends on to what vertical extent we integrate the gas density, which is uncertain beyond the scope of this work. For simplicity, we use the drift-limited $\tstop\Omega$ derived here and in the previous studies \citep[][]{Birnstiel2012,Okuzumi2012}.}
\begin{equation}
\eta=\frac{H^2}{2r^2}\left(\frac{7}{4}+q+\frac{r}{100\mathrm{au}}\right)\simeq 2\times 10^{-3}\left(\frac{9}{4}+\frac{r}{100\mathrm{au}}\right)\left(\frac{r}{100\mathrm{au}}\right)^{\frac{1}{2}},\label{eq:etab}
\end{equation}
where we use $q=1/2$. We approximate the radial drift velocity with $v_{\drift}\simeq-2\eta r\Omega\tstop\Omega$ and obtain
\begin{equation}
t_{\drift}\Omega\simeq\frac{1}{2\eta\tstop\Omega}.\label{eq:tdriftb}
\end{equation}

According to \citet[][]{Okuzumi2012}, we can expect that dust grains grow in size without significant radial drift when $t_{\mathrm{grow}}\lesssim t_{\drift}/10$ is satisfied\footnote{In \citet[][]{Okuzumi2012}, $t_{\mathrm{grow}}$ denotes the mass doubling timescale, and thus the coefficient of $t_{\drift}$ is different by a factor of three.}. Thus, using Equations (\ref{eq:tgrowEp}), (\ref{eq:etab}) and (\ref{eq:tdriftb}), we obtain the drift-limited stopping time
\begin{equation}
\tstop\Omega\simeq 0.13\left(\frac{\Sigma_{\dst,\mathrm{tot}}/\Sigma_{\mathrm{tot}}}{0.01}\right)\left(\frac{\eta}{4\times10^{-3}}\right)^{-1}.
\end{equation}
Our assumption $\tstop\Omega=0.6$ for $\Sigma_{\dst,\mathrm{tot}}/\Sigma_{\mathrm{gas}}=0.05$ is almost consistent with the above value in $30\;\mathrm{au}\leq r\leq 100\;\mathrm{au}$.  

\section{Dispersion relations of the secular GI for various power-law indices}\label{ap:qdepen}

In Section \ref{subsec:linear}, we conclude that the growth rate of the secular GI does not significantly depend on the dust drift. Plotting the dispersion relations for various power-law indices of the gas surface density ($q$), we show that this property is independent from the assumed $q$.

\begin{figure*}
	\begin{tabular}{c}
		\begin{minipage}{0.5\hsize}
			\begin{center}
				\includegraphics[width=0.9\columnwidth]{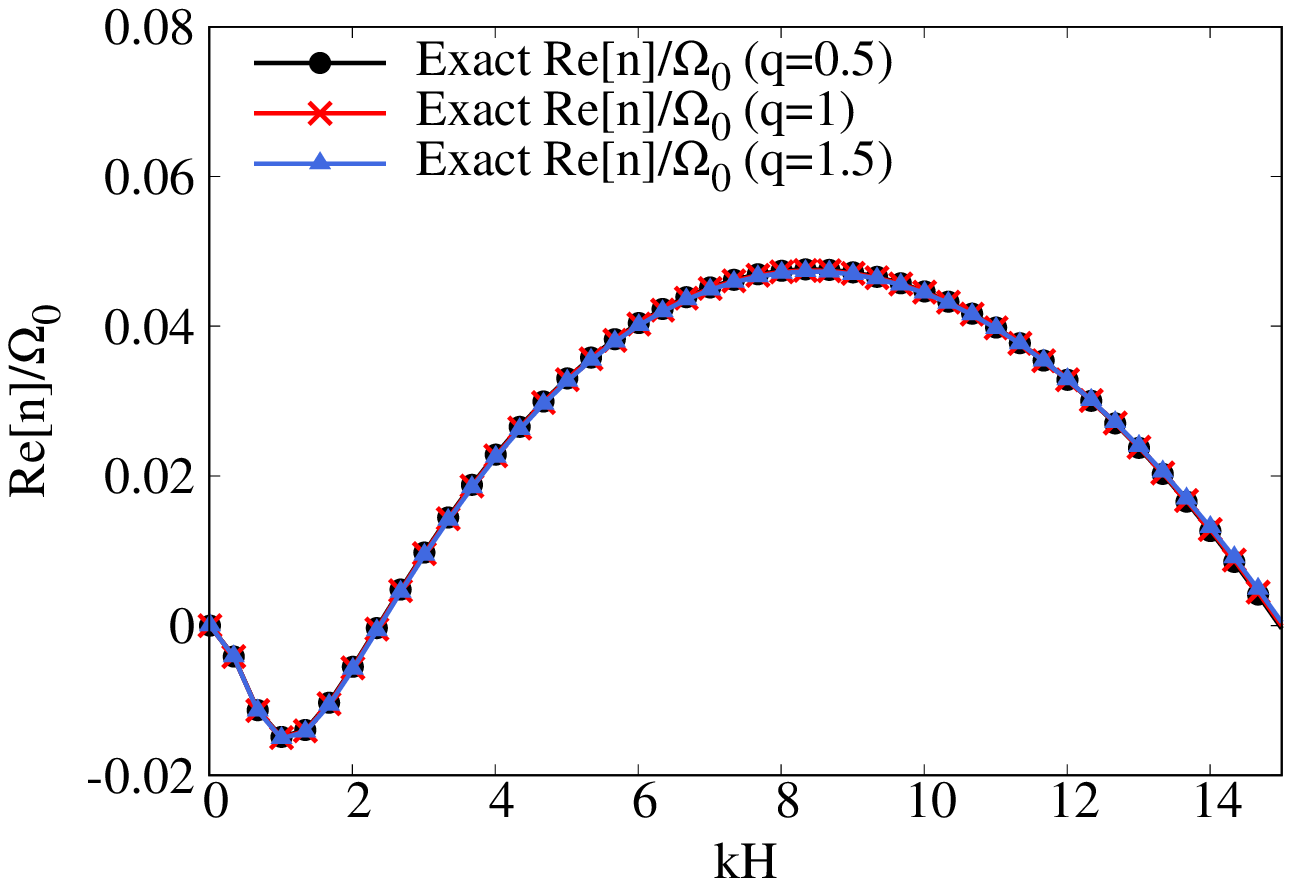}
			\end{center}
		\end{minipage}
		\begin{minipage}{0.5\hsize}
			\begin{center}
				\includegraphics[width=0.9\columnwidth]{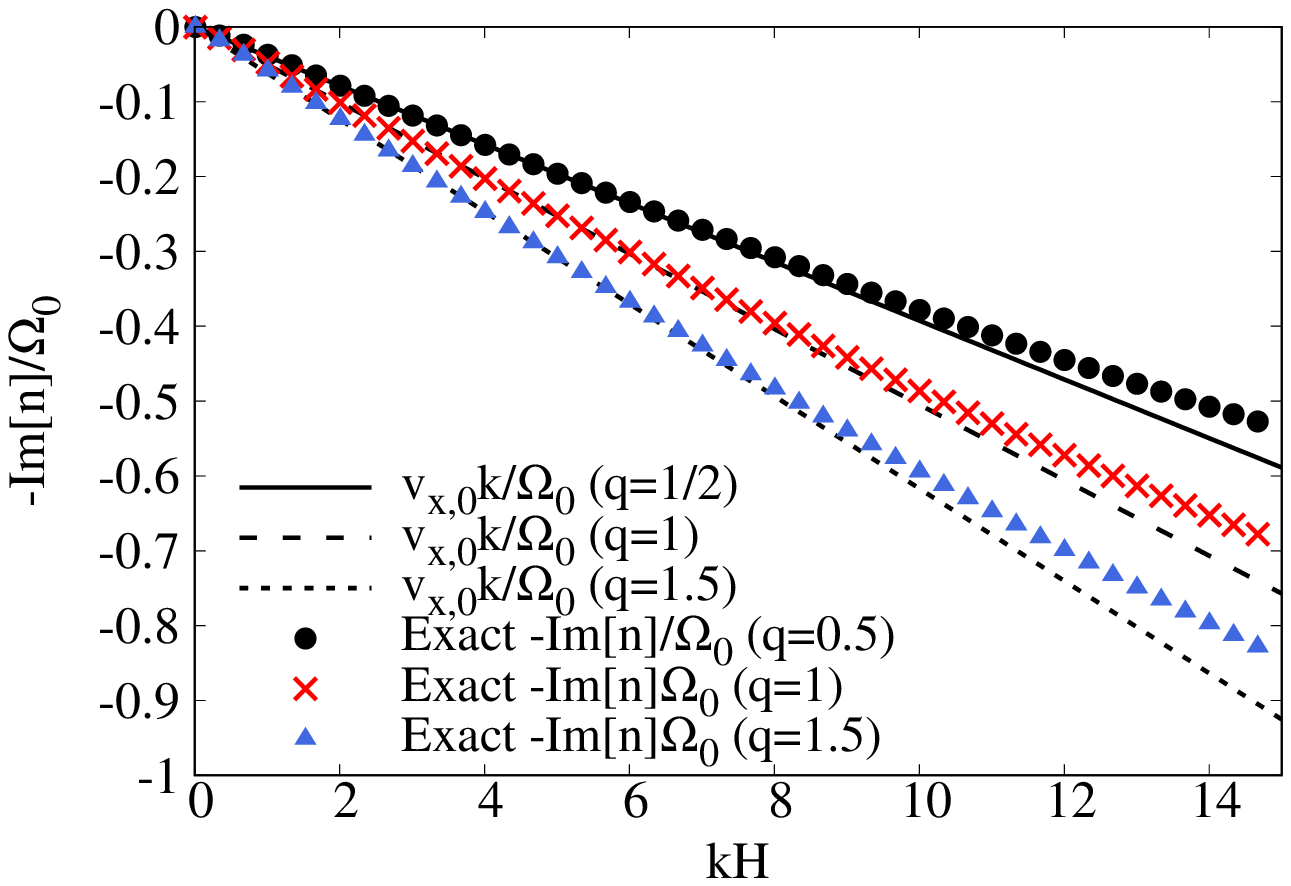}
			\end{center}
		\end{minipage}
	\end{tabular}
\caption{Dispersion relations of the secular GI without the turbulent viscosity for various power-law indices. We assume $\tstop\Omega=0.6$, $\Sigma_{\dst,0}/\Sigma_0=0.1$ and $\alpha=1\times10^{-3}$. The Toomre's $Q$ value is fixed to be $4.463$, which corresponds to the value at $r=75$ au in Q4a10 run ($q=0.5$). The horizontal and vertical axes are the same as in Figure \ref{fig:disp_Q3a5run}. The filled circles, the red cross symbols and blue triangles show dispersion relations for $q=0.5,\;1,\;1.5$, respectively. The solid, dashed and short dashed lines on the right panel show $v_{x,0}k/\Omega_0$ for $q=0.5,\;1,\;1.5$, respectively.}
 \label{fig:disp_variousq}
\end{figure*}

Figure \ref{fig:disp_variousq} shows the dispersion relations of the secular GI for $q=0.5,\;1,\;1.5$. We do not consider the turbulent viscosity in plotting Figure \ref{fig:disp_variousq} since the secular GI does not require the viscosity. 
We find that changing the value of $q$ does not significantly change the growth rate (see the left panel of Figure \ref{fig:disp_variousq}).
The oscillation frequencies are well captured by $v_{x,0}k/\Omega_0$ for each $q$ at small wavenumbers (see the right panel of Figure \ref{fig:disp_variousq}). The deviation from $v_{x,0}k/\Omega_0$ at large wavenumbers is due to the dust diffusion. Thus, regardless of the variety of the surface density distribution, the dust drift insignificantly change the growth rate of the secular GI.

\begin{figure*}
	\begin{tabular}{c}
		\begin{minipage}{0.5\hsize}
			\begin{center}
				\includegraphics[width=0.9\columnwidth]{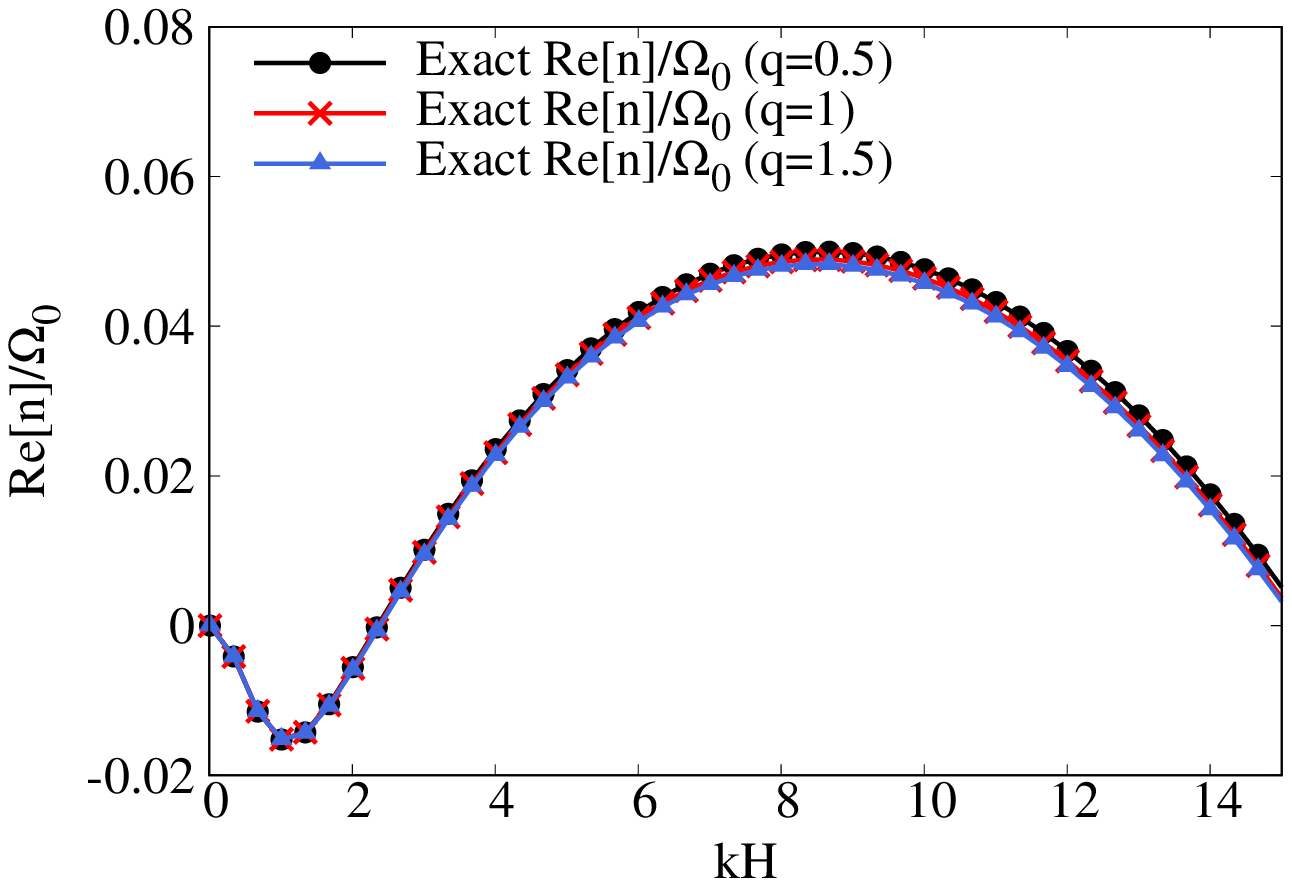}
			\end{center}
		\end{minipage}
		\begin{minipage}{0.5\hsize}
			\begin{center}
				\includegraphics[width=0.9\columnwidth]{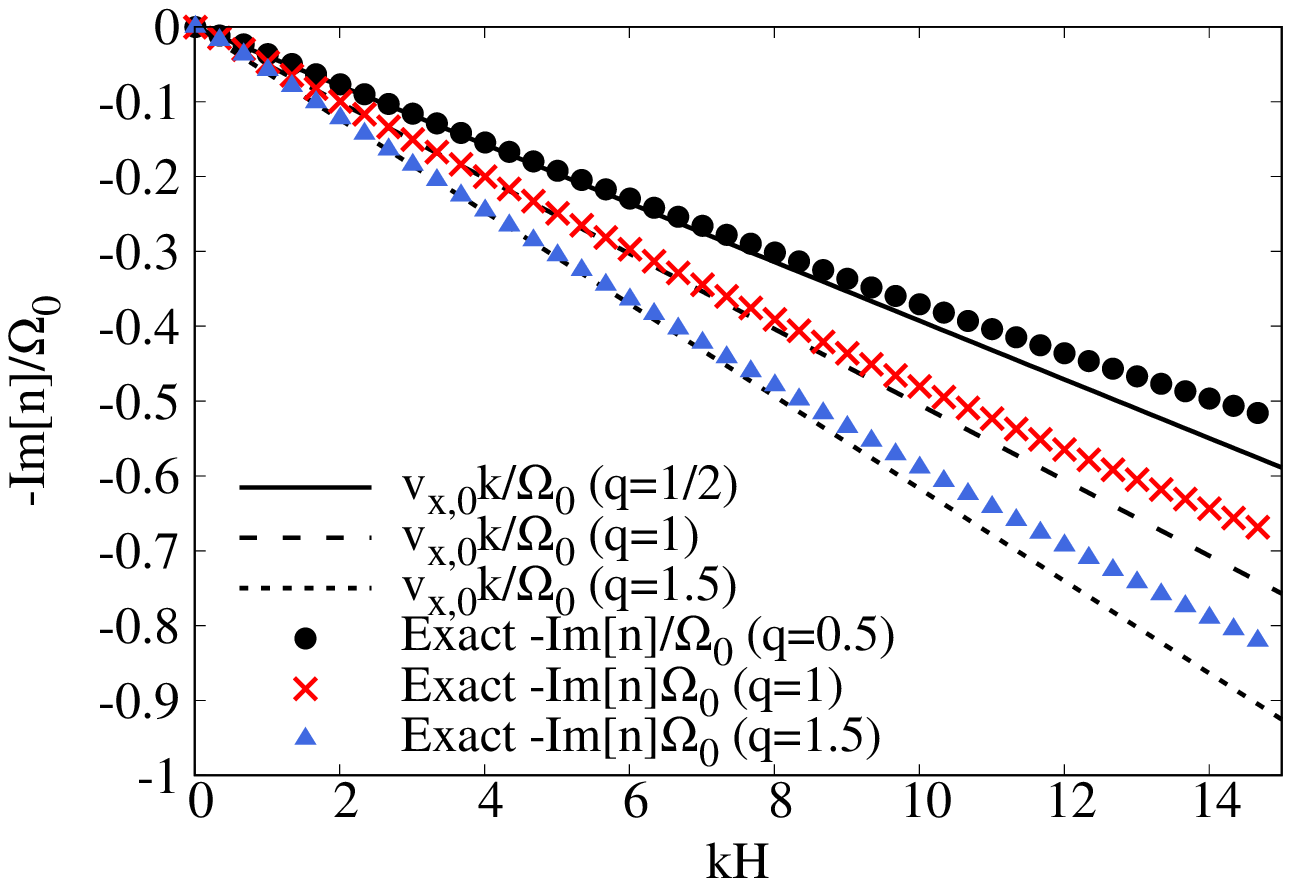}
			\end{center}
		\end{minipage}
	\end{tabular}
\caption{Dispersion relations of the secular GI with the turbulent viscosity for various power-law indices. We assume $\tstop\Omega=0.6$, $\Sigma_{\dst,0}/\Sigma_0=0.1$ and $\alpha=1\times10^{-3}$. The Toomre's $Q$ value is fixed to be $4.463$ as in Figure \ref{fig:disp_variousq}. The horizontal and vertical axes are the same as in Figures \ref{fig:disp_Q3a5run} and \ref{fig:disp_variousq}. The filled circles, the red cross symbols and blue triangles show dispersion relations for $q=0.5,\;1,\;1.5$, respectively. The solid, dashed and short dashed lines on the right panel show $v_{x,0}k/\Omega_0$ for $q=0.5,\;1,\;1.5$, respectively.}
 \label{fig:disp_variousq_w_vis}
\end{figure*}

When one considers the turbulent viscosity, the growth rate slightly changes depending on the assumed $q$. Figure \ref{fig:disp_variousq_w_vis} shows how the power-law index affects the dispersion relation in the presence of the turbulent viscosity. Although the growth rate is slightly smaller for larger $q$, the qualitative behavior is the same: the long-wavelength perturbations are stabilized by the Coriolis force, and the secular GI can be operational at intermediate wavelengths. The oscillation frequency is well described by $v_{x,0}k/\Omega_0$ for each $q$ in the presence of the small viscosity (e.g., $\alpha\lesssim10^{-3}$).




\bibliographystyle{aasjournal}
\bibliography{rttominaga2020}

\begin{thebibliography}{}
\expandafter\ifx\csname natexlab\endcsname\relax\def\natexlab#1{#1}\fi

\bibitem[{{Abod} {et~al.}(2019){Abod}, {Simon}, {Li}, {Armitage}, {Youdin}, \&
  {Kretke}}]{Abod2019}
{Abod}, C.~P., {Simon}, J.~B., {Li}, R., {et~al.} 2019, \apj, 883, 192

\bibitem[{{Alexiades} {et~al.}(1996){Alexiades}, {Amiez}, \&
  {Gremaud}}]{Alexiades1996}
{Alexiades}, V., {Amiez}, G., \& {Gremaud}, P.~A. 1996, CNME, 12, 31

\bibitem[{{ALMA Partnership} {et~al.}(2015){ALMA Partnership}, {Brogan},
  {P{\'e}rez}, {Hunter}, {Dent}, {Hales}, {Hills}, {Corder}, {Fomalont},
  {Vlahakis}, {Asaki}, {Barkats}, {Hirota}, {Hodge}, {Impellizzeri}, {Kneissl},
  {Liuzzo}, {Lucas}, {Marcelino}, {Matsushita}, {Nakanishi}, {Phillips},
  {Richards}, {Toledo}, {Aladro}, {Broguiere}, {Cortes}, {Cortes}, {Espada},
  {Galarza}, {Garcia-Appadoo}, {Guzman-Ramirez}, {Humphreys}, {Jung}, {Kameno},
  {Laing}, {Leon}, {Marconi}, {Mignano}, {Nikolic}, {Nyman}, {Radiszcz},
  {Remijan}, {Rod{\'o}n}, {Sawada}, {Takahashi}, {Tilanus}, {Vila Vilaro},
  {Watson}, {Wiklind}, {Akiyama}, {Chapillon}, {de Gregorio-Monsalvo}, {Di
  Francesco}, {Gueth}, {Kawamura}, {Lee}, {Nguyen Luong}, {Mangum}, {Pietu},
  {Sanhueza}, {Saigo}, {Takakuwa}, {Ubach}, {van Kempen}, {Wootten},
  {Castro-Carrizo}, {Francke}, {Gallardo}, {Garcia}, {Gonzalez}, {Hill},
  {Kaminski}, {Kurono}, {Liu}, {Lopez}, {Morales}, {Plarre}, {Schieven},
  {Testi}, {Videla}, {Villard}, {Andreani}, {Hibbard}, \&
  {Tatematsu}}]{ALMA-Partnership2015}
{ALMA Partnership}, {Brogan}, C.~L., {P{\'e}rez}, L.~M., {et~al.} 2015, \apjl,
  808, L3

\bibitem[{{Andrews} {et~al.}(2016){Andrews}, {Wilner}, {Zhu}, {Birnstiel},
  {Carpenter}, {P{\'e}rez}, {Bai}, {{\"O}berg}, {Hughes}, {Isella}, \&
  {Ricci}}]{Andrews2016}
{Andrews}, S.~M., {Wilner}, D.~J., {Zhu}, Z., {et~al.} 2016, \apjl, 820, L40

\bibitem[{{Andrews} {et~al.}(2018){Andrews}, {Huang}, {P{\'e}rez}, {Isella},
  {Dullemond}, {Kurtovic}, {Guzm{\'a}n}, {Carpenter}, {Wilner}, {Zhang}, {Zhu},
  {Birnstiel}, {Bai}, {Benisty}, {Hughes}, {{\"O}berg}, \&
  {Ricci}}]{Andrews2018}
{Andrews}, S.~M., {Huang}, J., {P{\'e}rez}, L.~M., {et~al.} 2018, \apjl, 869,
  L41

\bibitem[{{Bai} \& {Stone}(2010{\natexlab{a}})}]{Bai2010a}
{Bai}, X.-N., \& {Stone}, J.~M. 2010{\natexlab{a}}, \apj, 722, 1437

\bibitem[{{Bai} \& {Stone}(2010{\natexlab{b}})}]{Bai2010b}
---. 2010{\natexlab{b}}, \apjs, 190, 297

\bibitem[{{Bai} \& {Stone}(2010{\natexlab{c}})}]{Bai2010c}
---. 2010{\natexlab{c}}, \apjl, 722, L220

\bibitem[{{Birnstiel} {et~al.}(2009){Birnstiel}, {Dullemond}, \&
  {Brauer}}]{Birnstiel2009}
{Birnstiel}, T., {Dullemond}, C.~P., \& {Brauer}, F. 2009, \aap, 503, L5

\bibitem[{{Birnstiel} {et~al.}(2012){Birnstiel}, {Klahr}, \&
  {Ercolano}}]{Birnstiel2012}
{Birnstiel}, T., {Klahr}, H., \& {Ercolano}, B. 2012, \aap, 539, A148

\bibitem[{{Bitsch} \& {Johansen}(2017)}]{Bitsch2017}
{Bitsch}, B., \& {Johansen}, A. 2017, Astrophysics and Space Science Library,
  Vol. 445, {Planet Population Synthesis via Pebble Accretion}, ed. M.~{Pessah}
  \& O.~{Gressel}, 339

\bibitem[{{Br{\"u}gger} {et~al.}(2018){Br{\"u}gger}, {Alibert}, {Ataiee}, \&
  {Benz}}]{Brugger2018}
{Br{\"u}gger}, N., {Alibert}, Y., {Ataiee}, S., \& {Benz}, W. 2018, \aap, 619,
  A174

\bibitem[{{Carballido} {et~al.}(2006){Carballido}, {Fromang}, \&
  {Papaloizou}}]{Carballido2006}
{Carballido}, A., {Fromang}, S., \& {Papaloizou}, J. 2006, \mnras, 373, 1633

\bibitem[{{Chambers}(2018)}]{Chambers2018}
{Chambers}, J. 2018, \apj, 865, 30

\bibitem[{{Chen} \& {Lin}(2020)}]{Chen2020}
{Chen}, K., \& {Lin}, M.-K. 2020, \apj, 891, 132

\bibitem[{{Chiang} \& {Goldreich}(1997)}]{Chiang1997}
{Chiang}, E.~I., \& {Goldreich}, P. 1997, \apj, 490, 368

\bibitem[{{Clarke} {et~al.}(2018){Clarke}, {Tazzari}, {Juhasz}, {Rosotti},
  {Booth}, {Facchini}, {Ilee}, {Johns-Krull}, {Kama}, {Meru}, \&
  {Prato}}]{Clarke2018}
{Clarke}, C.~J., {Tazzari}, M., {Juhasz}, A., {et~al.} 2018, \apjl, 866, L6

\bibitem[{{Dipierro} {et~al.}(2018){Dipierro}, {Ricci}, {P{\'e}rez}, {Lodato},
  {Alexander}, {Laibe}, {Andrews}, {Carpenter}, {Chandler}, {Greaves}, {Hall},
  {Henning}, {Kwon}, {Linz}, {Mundy}, {Sargent}, {Tazzari}, {Testi}, \&
  {Wilner}}]{Dipierro2018}
{Dipierro}, G., {Ricci}, L., {P{\'e}rez}, L., {et~al.} 2018, \mnras, 475, 5296

\bibitem[{{Dubrulle} {et~al.}(1995){Dubrulle}, {Morfill}, \&
  {Sterzik}}]{Dubrulle1995}
{Dubrulle}, B., {Morfill}, G., \& {Sterzik}, M. 1995, \icarus, 114, 237

\bibitem[{{Dullemond} \& {Penzlin}(2018)}]{DullemondPenzlin2018}
{Dullemond}, C.~P., \& {Penzlin}, A.~B.~T. 2018, \aap, 609, A50

\bibitem[{{Dullemond} {et~al.}(2018){Dullemond}, {Birnstiel}, {Huang},
  {Kurtovic}, {Andrews}, {Guzm{\'a}n}, {P{\'e}rez}, {Isella}, {Zhu}, {Benisty},
  {Wilner}, {Bai}, {Carpenter}, {Zhang}, \& {Ricci}}]{Dullemond2018}
{Dullemond}, C.~P., {Birnstiel}, T., {Huang}, J., {et~al.} 2018, \apjl, 869,
  L46

\bibitem[{{Fedele} {et~al.}(2017){Fedele}, {Carney}, {Hogerheijde}, {Walsh},
  {Miotello}, {Klaassen}, {Bruderer}, {Henning}, \& {van
  Dishoeck}}]{Fedele2017}
{Fedele}, D., {Carney}, M., {Hogerheijde}, M.~R., {et~al.} 2017, \aap, 600, A72

\bibitem[{{Flock} {et~al.}(2015){Flock}, {Ruge}, {Dzyurkevich}, {Henning},
  {Klahr}, \& {Wolf}}]{Flock2015}
{Flock}, M., {Ruge}, J.~P., {Dzyurkevich}, N., {et~al.} 2015, \aap, 574, A68

\bibitem[{{Gerbig} {et~al.}(2020){Gerbig}, {Murray-Clay}, {Klahr}, \&
  {Baehr}}]{Gerbig2020}
{Gerbig}, K., {Murray-Clay}, R.~A., {Klahr}, H., \& {Baehr}, H. 2020, \apj,
  895, 91

\bibitem[{{Gole} {et~al.}(2020){Gole}, {Simon}, {Li}, {Youdin}, \&
  {Armitage}}]{Gole2020}
{Gole}, D.~A., {Simon}, J.~B., {Li}, R., {Youdin}, A.~N., \& {Armitage}, P.~J.
  2020, arXiv e-prints, arXiv:2001.10000

\bibitem[{{Gonzalez} {et~al.}(2015){Gonzalez}, {Laibe}, {Maddison}, {Pinte}, \&
  {M{\'e}nard}}]{Gonzalez2015}
{Gonzalez}, J.~F., {Laibe}, G., {Maddison}, S.~T., {Pinte}, C., \&
  {M{\'e}nard}, F. 2015, \mnras, 454, L36

\bibitem[{{Goodman} \& {Pindor}(2000)}]{Goodman2000}
{Goodman}, J., \& {Pindor}, B. 2000, \icarus, 148, 537

\bibitem[{{Hayashi}(1981)}]{Hayashi1981}
{Hayashi}, C. 1981, Progress of Theoretical Physics Supplement, 70, 35

\bibitem[{{Hu} {et~al.}(2019){Hu}, {Zhu}, {Okuzumi}, {Bai}, {Wang}, {Tomida},
  \& {Stone}}]{Hu2019}
{Hu}, X., {Zhu}, Z., {Okuzumi}, S., {et~al.} 2019, \apj, 885, 36

\bibitem[{{Huang} {et~al.}(2018){Huang}, {Andrews}, {Dullemond}, {Isella},
  {P{\'e}rez}, {Guzm{\'a}n}, {{\"O}berg}, {Zhu}, {Zhang}, {Bai}, {Benisty},
  {Birnstiel}, {Carpenter}, {Hughes}, {Ricci}, {Weaver}, \&
  {Wilner}}]{Huang2018}
{Huang}, J., {Andrews}, S.~M., {Dullemond}, C.~P., {et~al.} 2018, \apjl, 869,
  L42

\bibitem[{{Ilgner} \& {Nelson}(2006)}]{Ilgner2006}
{Ilgner}, M., \& {Nelson}, R.~P. 2006, \aap, 445, 205

\bibitem[{{Inoue} \& {Inutsuka}(2008)}]{Inoue2008}
{Inoue}, T., \& {Inutsuka}, S.-i. 2008, \apj, 687, 303

\bibitem[{{Isella} {et~al.}(2016){Isella}, {Guidi}, {Testi}, {Liu}, {Li}, {Li},
  {Weaver}, {Boehler}, {Carperter}, {De Gregorio-Monsalvo}, {Manara}, {Natta},
  {P{\'e}rez}, {Ricci}, {Sargent}, {Tazzari}, \& {Turner}}]{Isella2016}
{Isella}, A., {Guidi}, G., {Testi}, L., {et~al.} 2016, Physical Review Letters,
  117, 251101

\bibitem[{{Jacquet} {et~al.}(2011){Jacquet}, {Balbus}, \&
  {Latter}}]{Jacquet2011}
{Jacquet}, E., {Balbus}, S., \& {Latter}, H. 2011, \mnras, 415, 3591

\bibitem[{{Johansen} {et~al.}(2007){Johansen}, {Oishi}, {Mac Low}, {Klahr},
  {Henning}, \& {Youdin}}]{Johansen2007nature}
{Johansen}, A., {Oishi}, J.~S., {Mac Low}, M.-M., {et~al.} 2007, \nat, 448,
  1022

\bibitem[{{Johansen} \& {Youdin}(2007)}]{Johansen2007}
{Johansen}, A., \& {Youdin}, A. 2007, \apj, 662, 627

\bibitem[{{Johansen} {et~al.}(2009){Johansen}, {Youdin}, \& {Mac
  Low}}]{Johansen2009}
{Johansen}, A., {Youdin}, A., \& {Mac Low}, M.-M. 2009, \apjl, 704, L75

\bibitem[{{Kanagawa} {et~al.}(2015){Kanagawa}, {Muto}, {Tanaka}, {Tanigawa},
  {Takeuchi}, {Tsukagoshi}, \& {Momose}}]{Kanagawa2015}
{Kanagawa}, K.~D., {Muto}, T., {Tanaka}, H., {et~al.} 2015, \apjl, 806, L15

\bibitem[{{Kitamura} {et~al.}(2002){Kitamura}, {Momose}, {Yokogawa}, {Kawabe},
  {Tamura}, \& {Ida}}]{Kitamura2002}
{Kitamura}, Y., {Momose}, M., {Yokogawa}, S., {et~al.} 2002, \apj, 581, 357

\bibitem[{{Klahr} \& {Schreiber}(2020)}]{Klahr2020}
{Klahr}, H., \& {Schreiber}, A. 2020, arXiv e-prints, arXiv:2007.10696

\bibitem[{{Kobayashi} {et~al.}(2011){Kobayashi}, {Tanaka}, \&
  {Krivov}}]{Kobayashi2011}
{Kobayashi}, H., {Tanaka}, H., \& {Krivov}, A.~V. 2011, \apj, 738, 35

\bibitem[{{Kobayashi} {et~al.}(2010){Kobayashi}, {Tanaka}, {Krivov}, \&
  {Inaba}}]{Kobayashi2010}
{Kobayashi}, H., {Tanaka}, H., {Krivov}, A.~V., \& {Inaba}, S. 2010, \icarus,
  209, 836

\bibitem[{{Latter} \& {Rosca}(2017)}]{Latter2017}
{Latter}, H.~N., \& {Rosca}, R. 2017, \mnras, 464, 1923

\bibitem[{{Long} {et~al.}(2018){Long}, {Pinilla}, {Herczeg}, {Harsono},
  {Dipierro}, {Pascucci}, {Hendler}, {Tazzari}, {Ragusa}, {Salyk}, {Edwards},
  {Lodato}, {van de Plas}, {Johnstone}, {Liu}, {Boehler}, {Cabrit}, {Manara},
  {Menard}, {Mulders}, {Nisini}, {Fischer}, {Rigliaco}, {Banzatti}, {Avenhaus},
  \& {Gully-Santiago}}]{Long2018}
{Long}, F., {Pinilla}, P., {Herczeg}, G.~J., {et~al.} 2018, \apj, 869, 17

\bibitem[{{Michikoshi} {et~al.}(2012){Michikoshi}, {Kokubo}, \&
  {Inutsuka}}]{Michikoshi2012}
{Michikoshi}, S., {Kokubo}, E., \& {Inutsuka}, S.-i. 2012, \apj, 746, 35

\bibitem[{{Mizuno}(1980)}]{Mizuno1980}
{Mizuno}, H. 1980, Progress of Theoretical Physics, 64, 544

\bibitem[{{Nakagawa} {et~al.}(1986){Nakagawa}, {Sekiya}, \&
  {Hayashi}}]{Nakagawa1986}
{Nakagawa}, Y., {Sekiya}, M., \& {Hayashi}, C. 1986, \icarus, 67, 375

\bibitem[{{Ndugu} {et~al.}(2019){Ndugu}, {Bitsch}, \& {Jurua}}]{Ndugu2019}
{Ndugu}, N., {Bitsch}, B., \& {Jurua}, E. 2019, \mnras, 488, 3625

\bibitem[{{Okuzumi} {et~al.}(2016){Okuzumi}, {Momose}, {Sirono}, {Kobayashi},
  \& {Tanaka}}]{Okuzumi2016}
{Okuzumi}, S., {Momose}, M., {Sirono}, S.-i., {Kobayashi}, H., \& {Tanaka}, H.
  2016, \apj, 821, 82

\bibitem[{{Okuzumi} {et~al.}(2012){Okuzumi}, {Tanaka}, {Kobayashi}, \&
  {Wada}}]{Okuzumi2012}
{Okuzumi}, S., {Tanaka}, H., {Kobayashi}, H., \& {Wada}, K. 2012, \apj, 752,
  106

\bibitem[{{Ormel} \& {Cuzzi}(2007)}]{Ormel2007}
{Ormel}, C.~W., \& {Cuzzi}, J.~N. 2007, \aap, 466, 413

\bibitem[{{P{\'e}rez} {et~al.}(2019{\natexlab{a}}){P{\'e}rez}, {Casassus},
  {Baruteau}, {Dong}, {Hales}, \& {Cieza}}]{Perez2019a}
{P{\'e}rez}, S., {Casassus}, S., {Baruteau}, C., {et~al.} 2019{\natexlab{a}},
  \aj, 158, 15

\bibitem[{{P{\'e}rez} {et~al.}(2019{\natexlab{b}}){P{\'e}rez}, {Casassus},
  {Hales}, {Marino}, {Cheetham}, {Zurlo}, {Cieza}, {Dong}, {Alarc{\'o}n},
  {Ben{\'\i}tez-Llambay}, {Fomalont}, \& {Avenhaus}}]{Perez2019b}
{P{\'e}rez}, S., {Casassus}, S., {Hales}, A., {et~al.} 2019{\natexlab{b}},
  arXiv e-prints, arXiv:1906.06305

\bibitem[{{Pinte} {et~al.}(2018){Pinte}, {Price}, {M{\'e}nard}, {Duch{\^e}ne},
  {Dent}, {Hill}, {de Gregorio- Monsalvo}, {Hales}, \& {Mentiplay}}]{Pinte2018}
{Pinte}, C., {Price}, D.~J., {M{\'e}nard}, F., {et~al.} 2018, \apj, 860, L13

\bibitem[{{Pinte} {et~al.}(2019){Pinte}, {van der Plas}, {M{\'e}nard}, {Price},
  {Christiaens}, {Hill}, {Mentiplay}, {Ginski}, {Choquet}, {Boehler},
  {Duch{\^e}ne}, {Perez}, \& {Casassus}}]{Pinte2019}
{Pinte}, C., {van der Plas}, G., {M{\'e}nard}, F., {et~al.} 2019, Nature
  Astronomy, 3, 1109

\bibitem[{{Pinte} {et~al.}(2020){Pinte}, {Price}, {Menard}, {Duchene},
  {Christiaens}, {Andrews}, {Huang}, {Hill}, {van der Plas}, {Perez}, {Isella},
  {Boehler}, {Dent}, {Mentiplay}, \& {Loomis}}]{Pinte2020}
{Pinte}, C., {Price}, D.~J., {Menard}, F., {et~al.} 2020, arXiv e-prints,
  arXiv:2001.07720

\bibitem[{{Pollack} {et~al.}(1996){Pollack}, {Hubickyj}, {Bodenheimer},
  {Lissauer}, {Podolak}, \& {Greenzweig}}]{Pollack1996}
{Pollack}, J.~B., {Hubickyj}, O., {Bodenheimer}, P., {et~al.} 1996, \icarus,
  124, 62

\bibitem[{{Riols} \& {Lesur}(2019)}]{Riols2019}
{Riols}, A., \& {Lesur}, G. 2019, \aap, 625, A108

\bibitem[{{Sano} {et~al.}(2000){Sano}, {Miyama}, {Umebayashi}, \&
  {Nakano}}]{Sano2000}
{Sano}, T., {Miyama}, S.~M., {Umebayashi}, T., \& {Nakano}, T. 2000, \apj, 543,
  486

\bibitem[{{Sch{\"a}fer} {et~al.}(2020){Sch{\"a}fer}, {Johansen}, \&
  {Banerjee}}]{Schafer2020}
{Sch{\"a}fer}, U., {Johansen}, A., \& {Banerjee}, R. 2020, \aap, 635, A190

\bibitem[{{Schreiber} \& {Klahr}(2018)}]{Schreiber2018}
{Schreiber}, A., \& {Klahr}, H. 2018, \apj, 861, 47

\bibitem[{{Shadmehri}(2016)}]{Shadmehri2016a}
{Shadmehri}, M. 2016, \apj, 817, 140

\bibitem[{{Shadmehri} {et~al.}(2016){Shadmehri}, {Oudi}, \&
  {Rastegarzade}}]{Shadmehri2016b}
{Shadmehri}, M., {Oudi}, R., \& {Rastegarzade}, G. 2016, Research in Astronomy
  and Astrophysics, 16, 134

\bibitem[{{Shakura} \& {Sunyaev}(1973)}]{Shakura1973}
{Shakura}, N.~I., \& {Sunyaev}, R.~A. 1973, \aap, 24, 337

\bibitem[{{Shariff} \& {Cuzzi}(2011)}]{Shariff2011}
{Shariff}, K., \& {Cuzzi}, J.~N. 2011, \apj, 738, 73

\bibitem[{{Sheehan} \& {Eisner}(2017)}]{Sheehan2017}
{Sheehan}, P.~D., \& {Eisner}, J.~A. 2017, \apjl, 840, L12

\bibitem[{{Sheehan} \& {Eisner}(2018)}]{Sheehan2018}
---. 2018, \apj, 857, 18

\bibitem[{{Shu}(1984)}]{Shu1984}
{Shu}, F.~H. 1984, in IAU Colloq. 75: Planetary Rings, ed. R.~{Greenberg} \&
  A.~{Brahic}, 513--561

\bibitem[{{Simon} {et~al.}(2016){Simon}, {Armitage}, {Li}, \&
  {Youdin}}]{Simon2016}
{Simon}, J.~B., {Armitage}, P.~J., {Li}, R., \& {Youdin}, A.~N. 2016, \apj,
  822, 55

\bibitem[{{Stammler} {et~al.}(2019){Stammler}, {Dr{\c{a}}{\.z}kowska},
  {Birnstiel}, {Klahr}, {Dullemond}, \& {Andrews}}]{Stammler2019}
{Stammler}, S.~M., {Dr{\c{a}}{\.z}kowska}, J., {Birnstiel}, T., {et~al.} 2019,
  \apjl, 884, L5

\bibitem[{{Suriano} {et~al.}(2018){Suriano}, {Li}, {Krasnopolsky}, \&
  {Shang}}]{Suriano2018}
{Suriano}, S.~S., {Li}, Z.-Y., {Krasnopolsky}, R., \& {Shang}, H. 2018, \mnras,
  477, 1239

\bibitem[{{Suriano} {et~al.}(2019){Suriano}, {Li}, {Krasnopolsky}, {Suzuki}, \&
  {Shang}}]{Suriano2019}
{Suriano}, S.~S., {Li}, Z.-Y., {Krasnopolsky}, R., {Suzuki}, T.~K., \& {Shang},
  H. 2019, \mnras, 484, 107

\bibitem[{{Takahashi} \& {Inutsuka}(2014)}]{Takahashi2014}
{Takahashi}, S.~Z., \& {Inutsuka}, S.-i. 2014, \apj, 794, 55

\bibitem[{{Takahashi} \& {Inutsuka}(2016)}]{Takahashi2016}
---. 2016, \aj, 152, 184

\bibitem[{{Teague} {et~al.}(2018){Teague}, {Bae}, {Bergin}, {Birnstiel}, \&
  {Foreman-Mackey}}]{Teague2018}
{Teague}, R., {Bae}, J., {Bergin}, E.~A., {Birnstiel}, T., \& {Foreman-Mackey},
  D. 2018, \apjl, 860, L12

\bibitem[{{Tominaga} {et~al.}(2018){Tominaga}, {Inutsuka}, \&
  {Takahashi}}]{Tominaga2018}
{Tominaga}, R.~T., {Inutsuka}, S.-i., \& {Takahashi}, S.~Z. 2018, \pasj, 70, 3

\bibitem[{{Tominaga} {et~al.}(2019){Tominaga}, {Takahashi}, \&
  {Inutsuka}}]{Tominaga2019}
{Tominaga}, R.~T., {Takahashi}, S.~Z., \& {Inutsuka}, S.-i. 2019, \apj, 881, 53

\bibitem[{{Tsukagoshi} {et~al.}(2016){Tsukagoshi}, {Nomura}, {Muto}, {Kawabe},
  {Ishimoto}, {Kanagawa}, {Okuzumi}, {Ida}, {Walsh}, \&
  {Millar}}]{Tsukagoshi2016}
{Tsukagoshi}, T., {Nomura}, H., {Muto}, T., {et~al.} 2016, \apjl, 829, L35

\bibitem[{{Tsukagoshi} {et~al.}(2019){Tsukagoshi}, {Momose}, {Kitamura},
  {Saito}, {Kawabe}, {Andrews}, {Wilner}, {Kudo}, {Hashimoto}, {Ohashi}, \&
  {Tamura}}]{Tsukagoshi2019}
{Tsukagoshi}, T., {Momose}, M., {Kitamura}, Y., {et~al.} 2019, \apj, 871, 5

\bibitem[{{Umurhan} {et~al.}(2020){Umurhan}, {Estrada}, \&
  {Cuzzi}}]{Umurhan2020}
{Umurhan}, O.~M., {Estrada}, P.~R., \& {Cuzzi}, J.~N. 2020, \apj, 895, 4

\bibitem[{{Vandervoort}(1970)}]{Vandervoort1970}
{Vandervoort}, P.~O. 1970, \apj, 161, 87

\bibitem[{{Ward}(2000)}]{Ward2000}
{Ward}, W.~R. 2000, {On Planetesimal Formation: The Role of Collective Particle
  Behavior}, ed. R.~M. {Canup}, K.~{Righter}, \& {et al.}, 75--84

\bibitem[{{Whipple}(1972)}]{Whipple1972}
{Whipple}, F.~L. 1972, in From Plasma to Planet, ed. A.~{Elvius}, 211

\bibitem[{{Yang} \& {Johansen}(2014)}]{Yang2014}
{Yang}, C.-C., \& {Johansen}, A. 2014, \apj, 792, 86

\bibitem[{{Yang} {et~al.}(2017){Yang}, {Johansen}, \& {Carrera}}]{Yang2017}
{Yang}, C.~C., {Johansen}, A., \& {Carrera}, D. 2017, \aap, 606, A80

\bibitem[{{Yang} {et~al.}(2018){Yang}, {Mac Low}, \& {Johansen}}]{Yang2018}
{Yang}, C.-C., {Mac Low}, M.-M., \& {Johansen}, A. 2018, \apj, 868, 27

\bibitem[{{Yang} \& {Zhu}(2020)}]{Yang2020}
{Yang}, C.-C., \& {Zhu}, Z. 2020, \mnras, 491, 4702

\bibitem[{{Youdin} \& {Johansen}(2007)}]{Youdin2007}
{Youdin}, A., \& {Johansen}, A. 2007, \apj, 662, 613

\bibitem[{{Youdin}(2005{\natexlab{a}})}]{Youdin2005a}
{Youdin}, A.~N. 2005{\natexlab{a}}, ArXiv Astrophysics e-prints,
  astro-ph/0508659

\bibitem[{{Youdin}(2005{\natexlab{b}})}]{Youdin2005b}
---. 2005{\natexlab{b}}, ArXiv Astrophysics e-prints, astro-ph/0508662

\bibitem[{{Youdin}(2011)}]{Youdin2011}
---. 2011, \apj, 731, 99

\bibitem[{{Youdin} \& {Goodman}(2005)}]{Youdin2005}
{Youdin}, A.~N., \& {Goodman}, J. 2005, \apj, 620, 459

\bibitem[{{Youdin} \& {Lithwick}(2007)}]{YL2007}
{Youdin}, A.~N., \& {Lithwick}, Y. 2007, \icarus, 192, 588

\bibitem[{{Zhang} {et~al.}(2015){Zhang}, {Blake}, \& {Bergin}}]{Zhang2015}
{Zhang}, K., {Blake}, G.~A., \& {Bergin}, E.~A. 2015, \apjl, 806, L7

\bibitem[{{Zhang} {et~al.}(2018){Zhang}, {Zhu}, {Huang}, {Guzm{\'a}n},
  {Andrews}, {Birnstiel}, {Dullemond}, {Carpenter}, {Isella}, {P{\'e}rez},
  {Benisty}, {Wilner}, {Baruteau}, {Bai}, \& {Ricci}}]{Zhang2018}
{Zhang}, S., {Zhu}, Z., {Huang}, J., {et~al.} 2018, \apjl, 869, L47

\bibitem[{{Zhu} {et~al.}(2019){Zhu}, {Zhang}, {Jiang}, {Kataoka}, {Birnstiel},
  {Dullemond}, {Andrews}, {Huang}, {P{\'e}rez}, {Carpenter}, {Bai}, {Wilner},
  \& {Ricci}}]{Zhu2019}
{Zhu}, Z., {Zhang}, S., {Jiang}, Y.-F., {et~al.} 2019, \apjl, 877, L18

\end{thebibliography}

%
%
%


\end{document}